\documentclass{aa}
\usepackage{graphicx}
\usepackage{txfonts}
\usepackage{natbib}
\usepackage{booktabs}
\usepackage{amsmath}
\usepackage{placeins}

\usepackage{hyperref}
\hypersetup{colorlinks=true,linkcolor=blue,citecolor=blue,urlcolor=blue,
 pdftitle={Unresolved binaries from Gaia XP to SPHEREx},
 pdfauthor={Jiadong Li}}
\bibpunct{(}{)}{;}{a}{}{,}

\newcommand{\Gaia}{\textit{Gaia}}
\newcommand{\SPHEREx}{\textit{SPHEREx}}
\newcommand{\teff}{\ensuremath{T_{\mathrm{eff}}}}
\newcommand{\mh}{\ensuremath{[\mathrm{M/H}]}}
\newcommand{\feh}{\ensuremath{[\mathrm{Fe/H}]}}
\newcommand{\Msun}{\ensuremath{M_\odot}}
\newcommand{\MKs}{\ensuremath{M_{K_s}}}
\newcommand{\GKs}{\ensuremath{G-K_s}}
\newcommand{\dchi}{\ensuremath{\Delta\chi^2}}
\newcommand{\mum}{\ensuremath{\,\mu\mathrm m}}

\journalname{Preprint}
\makeatletter
\renewcommand*\aa@manuscriptname{}
\renewcommand*\aa@textidlineempty{\aa@headings}
\makeatother

\begin{document}

\title{Unresolved binaries from \Gaia\ XP to \SPHEREx}
\titlerunning{Unresolved binaries from XP to SPHEREx}
\author{Jiadong Li\inst{1} \and Hans-Walter Rix\inst{1} \and Johanna M\"uller-Horn\inst{1} \and Kareem El-Badry\inst{2,1} \and Ben Pennell\inst{1} \and Xiaoying Pang\inst{3} \and Chao~Liu\inst{4,5}}
\authorrunning{J. Li et al.}
\institute{Max-Planck-Institut f\"ur Astronomie, K\"onigstuhl 17, D-69117 Heidelberg, Germany\\
 \email{jdli@mpia.de}
 \and
 Department of Astronomy, California Institute of Technology, 1200 E. California Blvd., Pasadena, CA 91125, USA
 \and
 Department of Physics, Xi'an Jiaotong-Liverpool University, Suzhou 215123, China
 \and
 Key Laboratory of Space Astronomy and Technology, National Astronomical Observatories, CAS, Beijing 100101, China
 \and
 University of Chinese Academy of Sciences, Beijing 100049, China}
\date{}

\abstract{Unresolved companions alter the luminosity and spectra of a star, but a colour--magnitude diagram (CMD) reveals them only when the system is over-luminous, and a \Gaia\ photocentre orbit constrains only a combination of the mass and flux ratios. We fit the absolute \Gaia\ BP/RP (XP) and \SPHEREx\ spectra (XPS) of about 71\,000 low-mass stars in the \Gaia\ Catalogue of Nearby Stars with the J-CaPS XPS emulator, which predicts single-star and coeval-binary spectra on an absolute flux scale.
First, more than half of known binaries lie within the main-sequence (MS) band of the CMD, because the companion is faint or lifts a metal-poor primary onto the single-star ridge. XPS recovers 86 per cent of the binaries that a CMD cut finds and 44 per cent of those it misses.
Second, a small photocentre motion can come from a faint companion or a luminous one of nearly equal mass. XPS distinguishes them and recovers the mass ratios of double-lined binaries. Among substellar candidates with radial-velocity follow-up, it ranks all 11 binaries above all 16 confirmed brown-dwarf or planet hosts. The public package \texttt{spbin} provides this check before follow-up.
Third, where XPS detects no companion light, we place an upper limit on the light of an MS companion. Among \Gaia\ DR3 astrometric orbits with clean spectral fits, 2841 require more companion mass than XPS allows for a single MS star. Within 200\,pc, their companion masses cluster at those of typical white dwarfs, and nearly circular orbits support this interpretation. 
Of 24 companions that joint radial-velocity and astrometric orbits show to be faint, XPS recovers 22, compared with 14 selected by the absence of a colour excess. Applied to forthcoming \Gaia\ DR4 orbital solutions, XPS can help distinguish luminous binaries from substellar companions and identify candidate compact companions whose dynamical masses exceed the limits allowed by their light.}
\maketitle

\section{Introduction}\label{sec:intro}

An unresolved companion adds light, brightening the system relative to the primary alone in the colour--magnitude diagram (CMD; \citealt{Li2025}). When its temperature differs from that of the primary, it changes the shape of the spectra, which biases the labels of a single-star fit \citep{ElBadry2018sig}. The \Gaia\ Apsis pipeline accounts for such composite spectra with coeval-binary models \citep{Creevey2023,Fouesneau2023}. \Gaia\ measures the orbit of the photocentre about the centre of mass \citep{Halbwachs2023,GaiaArenou2023}. Light from the companion shifts the photocentre towards it, reducing the measured orbital amplitude for a pair of main-sequence (MS) stars. Radial velocities (RVs) and excess astrometric noise \citep{Belokurov2020,Penoyre2020} detect a companion through its orbital motion, and imaging separates the two stars when they are far enough apart.

Large samples rarely measure the light of the companion itself. A star over-luminous in the CMD is counted as a binary, a photocentre orbit is converted into a companion mass, and an orbit too large for an MS companion is taken as evidence of a white dwarf, a neutron star or a black hole. Each of these inferences assumes how much light the companion contributes. When the assumption is wrong, the companion biases the inferred properties of the system: a single-star fit to the combined light biases the luminosity, mass, age and metallicity of the primary, and an orbit interpreted without the companion light biases its mass ratio. These biases carry over to the inferred frequency and mass-ratio distribution of companions, which link star formation to the present-day stellar population \citep{Duchene2013,Moe2017,Offner2023}. The solar neighbourhood, where the multiplicity census is most complete \citep{Duquennoy1991,Fischer1992,Raghavan2010,Janson2012,Tokovinin2014,WardDuong2015,Winters2019}, is also where the light of companions can be measured most directly.

Cluster members share a similar age and metallicity and form a narrow single-star sequence, on which the extra light of a companion stands out. Infrared photometry increases the sensitivity to cool companions, which contribute a larger fraction of the flux at longer wavelengths, and finds companions down to low mass ratios \citep{Malhotra2026}. In the field each signature alone is degenerate. Metallicity broadens the single-star sequence, and a metal-poor primary with a luminous companion can sit on it. A near-equal-mass pair moves the photocentre little, so the orbit of a luminous companion looks like that of a low-mass companion \citep{Halbwachs2023,BailerJones2026}. The same degeneracy limits the search for compact companions in \Gaia\ orbits. An orbit that needs more mass than a single MS companion with the observed light can supply points to a white dwarf, a neutron star or a black hole \citep{Shahaf2023,Shahaf2024,ElBadry2024NS}, but deciding how much light the companion may add requires either a mass--luminosity relation or a measured colour. Both are misled by an unresolved pair of M dwarfs, which is fainter than a single star of the same total mass and so imitates a compact object.

Absolute spectra measure the combined light of the pair directly, without a resolved orbit or a measurable velocity separation. Their reach is set by the contrast between the two stars and by how well a composite can be told apart from the range of single-star spectra. \citet{Li2025} showed the scale of this approach with \Gaia\ BP/RP spectra (hereafter XP): comparing data-driven single-star and binary spectra on an absolute flux scale, they identified about 14 million binary candidates within 1\,kpc, at mass ratios $0.5\le q\le1$. That range reflects two limits of optical data. At low mass ratios a cool companion adds little optical flux, and near equal mass the composite has the shape of a single star and stands out only by its doubled luminosity. The infrared addresses the first limit. When the companion is cooler than the primary, the flux ratio rises towards the infrared, and the molecular bands of a cool photosphere carry more information than a single infrared excess. \SPHEREx\ provides all-sky spectra between 0.75 and 5$\mum$ \citep{Bock2026}, which reach the stars of the \Gaia\ Catalogue of Nearby Stars \citep[GCNS;][]{Smart2021} and the cool, low-mass companions that add little light to XP.

In this paper we extend the method of \citet{Li2025} into the infrared. We fit the absolute XP and \SPHEREx\ spectra (together XPS) of about 71\,000 primaries of 0.1--0.9\,\Msun\ in the GCNS, measure the light of their unresolved companions, and compare it with what their CMD positions and \Gaia\ orbits imply. We address three questions. Which binaries does the CMD miss, and does XPS find them (Sect.~\ref{sec:gcns})? Can the spectra distinguish the faint and luminous solutions of a photocentre orbit (Sect.~\ref{sec:gcns-orbits})? And when no companion light is seen, which companions does that exclude, and which orbits then require a companion too faint for its mass (Sect.~\ref{sec:gcns-limits})? Sections~\ref{sec:data-sample} and \ref{sec:model} describe the data and the model, and Sect.~\ref{sec:discussion} discusses the limitations of the method.

\section{Data and samples}\label{sec:data-sample}

We fit the absolute spectra of nearby low-mass stars with the J-CaPS XPS emulator (hereafter the emulator; Sect.~\ref{sec:model}).

\subsection{Data vector}\label{sec:data}

We describe each star by its XPS spectra. The XP spectra are sampled at 61 wavelengths from 392 to 992\,nm in steps of 10\,nm (hereafter XP samples; \citealt{DeAngeli2023,Montegriffo2023}). The \SPHEREx\ spectra have 102 channels from 0.753 to 4.982$\mum$.\footnote{We extract the \SPHEREx\ spectra by point-spread-function photometry on the Quick Release 2 spectral images of the NASA/IPAC Infrared Science Archive (\url{https://doi.org/10.26131/IRSA652}) with the code \texttt{spherextract} (\url{https://github.com/freddavies/spherextract}).} The emulator also predicts 2MASS $J$, $H$, $K_s$ \citep{Skrutskie2006} and unWISE $W1$, $W2$ \citep{Schlafly2019} fluxes. 
The continuous fits of the GCNS stars include $JHK_s$, the grid fits of the benchmark sample do not, and neither includes $W1$ or $W2$.

The \SPHEREx\ channels have a resolving power of about 40 below 3.8$\mum$ and 110--130 above \citep{Bock2026}. For a typical star of the sample we fit about 100 channels, each with nearly twice the signal-to-noise ratio (S/N) of an XP sample. \SPHEREx\ overestimates the flux of bright sources, so we drop every channel brighter than AB magnitude 9. 
This channel cut still matters for our sample at $G\ge9$, because an M dwarf is much brighter in the infrared than in the $G$ band. All fluxes are $F_\lambda$ in units of $10^{-18}$\,W\,m$^{-2}$\,nm$^{-1}$, scaled to 10\,pc with the parallax $\varpi$,
\begin{equation}\label{eq:f10}
 F_{10}=F_{\mathrm{obs}}\left(\frac{100}{\varpi/\mathrm{mas}}\right)^2 ,
\end{equation}
with the same scaling applied to the uncertainties. All comparisons of single-star and binary models in this paper use these absolute fluxes.

\subsection{Stellar labels for training}\label{sec:labels}

To train the emulator, we compiled stellar labels from the literature (Table~\ref{tab:labels}). Stars hotter than 4000\,K take spectroscopic labels from the Apache Point Observatory Galactic Evolution Experiment (APOGEE) \citep{Majewski2017,Meszaros2025}. M dwarfs take their labels from one of six literature catalogues \citep{Birky2020,Schweitzer2019,Newton2015,Mann2015,Maldonado2020,Marfil2021}. A small set of ultracool dwarfs with evolutionary temperatures \citep{Sanghi2023} extends the network to the coolest dwarfs.

\begin{table}
\caption{Label sources of the data-driven network.}\label{tab:labels}
\centering
\small\setlength{\tabcolsep}{2.5pt}
\begin{tabular}{lrrcc}
\toprule
Source & Train & Valid. & $\teff$ [K] & $\feh$ \\
\midrule
APOGEE ASPCAP$^{a}$ & 5847 & 1462 & 4000--6963 & $-1.72$--$+0.42$ \\
\citet{Birky2020} & 605 & 129 & 2948--4017 & $-0.43$--$+0.37$ \\
\citet{Schweitzer2019} & 124 & 26 & 2964--4143 & $-0.34$--$+0.29$ \\
\citet{Newton2015} & 104 & 19 & 3200--4027 & $-0.71$--$+0.80$ \\
\citet{Mann2015} & 86 & 21 & 2800--4124 & $-0.54$--$+0.53$ \\
\citet{Maldonado2020} & 51 & 12 & 3257--3942 & $-0.40$--$+0.17$ \\
\citet{Marfil2021} & 26 & 6 & 3029--4049 & $-0.91$--$+0.15$ \\
UCDs, \citet{Sanghi2023}$^{b}$ & 24 & 7 & 2313--2929 & -- \\
\midrule
Total & 6867 & 1682 & 2313--6963 & $-1.72$--$+0.80$ \\
\bottomrule
\end{tabular}
\tablefoot{Ranges span training and validation stars. A star listed in several catalogues takes one label, in the order Mann, Schweitzer, Marfil, Maldonado, Newton, Birky. No zero point is fitted between catalogues, and Birky labels require the catalogue's safe flag. $^{a}$ Astra ASPCAP v0.7 dwarf labels (APOGEE Stellar Parameters and Chemical Abundances Pipeline) \citep{SDSSDR19,Meszaros2025}. $^{b}$ Evolutionary temperatures. Metallicity unknown, marginalised over $-0.5$ to $+0.5$.}
\end{table}

\subsection{The GCNS sample}\label{sec:gcns-sample}

The parent sample contains the 143\,523 GCNS sources with complete XPS data that were not used to train, validate or calibrate the emulator. For the analysis of the GCNS sample in Sects.~\ref{sec:gcns}--\ref{sec:gcns-limits}, we select the stars with $9\le G<17.7$ and a single-star primary mass of 0.1--0.9\,\Msun. We remove sources that the GCNS lists in a resolved wide binary, sources with a multi-peaked image ($\texttt{ipd\_frac\_multi\_peak}>0$), white-dwarf candidates (white-dwarf probability above 0.05) and sources more than 0.5\,mag below the MS ridge. We define the ridge as the median \MKs, in bins of $BP-RP$, of the stars without a \Gaia\ non-single-star solution. Stars with such a solution stay in the sample. This leaves 70\,933 stars.

\subsection{\Gaia\ DR3 orbits}\label{sec:orbits-data}

The orbits come from the non-single-star catalogue of \Gaia\ Data Release~3 (DR3) \citep{GaiaArenou2023,Halbwachs2023}. It contains two-body orbits fitted to the astrometry alone (solution type \texttt{Orbital}), to the astrometry and single-lined RVs together (\texttt{AstroSpectroSB1}), or to RVs alone for single-lined (SB1) or double-lined (SB2) spectroscopic binaries. For some of them, its \texttt{binary\_masses} table derives component masses, for example by combining an astrometric orbit with an SB2 solution (\texttt{Orbital+SB2}) or with an isochrone mass of the primary (\texttt{Orbital+M1}).

\subsection{Benchmark sample}\label{sec:sample}

To test the fits on real binaries, we select from the parent sample the binaries whose mass ratio is known independently of our fits. These are \Gaia\ DR3 orbital solutions with a primary mass, for both SB2 systems and astrometric SB1 systems \citep{GaiaArenou2023}, detached eclipsing binaries from DEBCat \citep{Southworth2015}, and APOGEE SB2 systems \citep{ElBadry2018}. For the \Gaia\ SB1 systems, the \texttt{binary\_masses} table combines the \texttt{AstroSpectroSB1} orbit with an isochrone mass of the primary. We require $G\ge8$, $\MKs\ge3$ and, where known, $M_1\le1\,\Msun$, which leaves 816 binaries. As controls, we pair each binary with the two field stars without a \Gaia\ orbital solution that are nearest to it in $G$ and $BP-RP$. We compare the binary detection statistic of each known binary with those of its matched controls.

\section{Methods}\label{sec:model}

A companion changes the spectra of a star in two ways. It adds light, so the pair is brighter than the primary alone, and when it is cooler than the primary it adds proportionally more light at long wavelengths. Comparing single-star and binary models on absolute fluxes, with the flux scale free only within the parallax uncertainty, keeps both signals. Normalising the spectra would remove the first. The model therefore has to predict the absolute 10\,pc spectra of a star of mass $M$, metallicity $\mh$ and age $\tau$, and the sum of two such spectra, for XPS.\footnote{J-CaPS, a \textbf{J}AX-based code inspired by The \textbf{Ca}nnon \citep{Ness2015}, The \textbf{P}ayne \citep{Ting2019} and \textbf{S}LAM \citep{Zhang2020SLAM}, is the stellar parameter fitting code of \citet{Li2025, Li2026}.} The approach follows \citet{Li2025}, who compared single-star and binary models on absolute XP spectra. Here we add the \SPHEREx\ spectra and extend the model to the lowest-mass stars and brown dwarfs. The emulator consists of a neural network that turns stellar labels into spectra (Sect.~\ref{sec:network}), an interface from mass, metallicity and age to those labels, with theoretical models where the network does not apply (Sect.~\ref{sec:routes}), and the sum of two components (Sect.~\ref{sec:binary}).

\subsection{A data-driven model of single-star spectra}\label{sec:network}

The network predicts spectra from four labels: effective temperature \teff, absolute magnitude in the 2MASS $K_s$ band \MKs, colour \GKs\ and metallicity \feh. Temperature, colour and metallicity set the shape of the spectra, and \MKs\ sets their absolute level,
\begin{equation}\label{eq:anchor}
 \log_{10}F_\lambda(10\,\mathrm{pc}) = r_\lambda - r_{K_s} + \log_{10}F_{K_s,0} - 0.4\,\MKs ,
\end{equation}
where $r_\lambda(\teff,\GKs,\feh)$ is the network output, $r_{K_s}$ its value in the $K_s$ band and $F_{K_s,0}$ the $K_s$ zero point. Tying the level to a measured magnitude leaves the network to learn only relative spectral shapes, and $K_s$ is also the band in which the mass--luminosity relation of M dwarfs is calibrated \citep{Mann2019}. The approach follows The Cannon \citep{Ness2015}, which is data-driven, and The Payne \citep{Ting2019}, which interpolates model spectra with a neural network. For XP spectra, \citet{Andrae2023} estimate stellar labels and \citet{Zhang2023} model the spectra generatively. The network is a small multilayer perceptron with two hidden layers of 16 units, trained on the stars of Sect.~\ref{sec:labels} with a robust loss on the linear fluxes as described in Appendix~\ref{app:training}.

\subsection{From mass, metallicity and age to spectra}\label{sec:routes}

The network takes \teff, \MKs, \GKs\ and \feh\ as input, whereas the fits need spectra as a function of mass, metallicity and age. For a given mass, metallicity and age, PARSEC isochrones \citep{Bressan2012,Chen2014} give \teff, \MKs\ and \GKs, and the network turns these labels into spectra wherever they lie inside its training domain. The data-driven model covers most stars above 0.09\,\Msun. In a narrow gap at 0.136--0.147\,\Msun\ the isochrone labels fall between the two groups of cool training stars, the M dwarfs with spectroscopic labels and the ultracool dwarfs with evolutionary temperatures, and the emulator interpolates the logarithmic flux of each channel linearly in mass. At lower masses it uses theoretical evolution and atmosphere models: ATMO\,2020 evolution and equilibrium-chemistry atmospheres \citep{Phillips2020} below 0.075\,\Msun, and BHAC15 evolution \citep{Baraffe2015} with BT-Settl atmospheres \citep{Allard2012} from there up to the data-driven model, in both cases with the atmosphere flux scaled by $(R/10\,\mathrm{pc})^2$ for the model radius $R$. At the high-mass end no spectra are returned above about 1\,\Msun\ at ages above 3\,Gyr, where the star begins to leave the MS and its labels leave those of the training dwarfs. The GCNS primaries of 0.1--0.9\,\Msun\ are therefore covered at all ages from 0.5 to 9.9\,Gyr. For the more massive primaries of some \Gaia\ DR3 orbits, up to 1.25\,\Msun, the fits are restricted to the ages at which the emulator returns spectra.

The data-driven model reproduces observed spectra and accepts any metallicity in its training range, whereas the interpolated gap and the theoretical models exist at solar metallicity only and are model predictions. In a binary model the difference matters mainly for the companion: the companions of the lowest-mass primaries come from the theoretical models at almost every $q$, those of a 0.5\,\Msun\ primary only at low $q$. Age matters mainly for a substellar companion. Stellar spectra change little between 0.5 and 9.9\,Gyr, but brown dwarfs never reach stable hydrogen burning and cool throughout their lives, so at fixed mass their flux and spectral shape depend strongly on age \citep{Burrows1997,Phillips2020}, and below the hydrogen-burning limit objects fade by more than an order of magnitude in the infrared between these ages.

\subsection{Unresolved binaries}\label{sec:binary}

We write the absolute spectra of a star of mass $M$ as $F_\lambda(M)\equiv F_\lambda(M,\mh,\tau)$ and model an unresolved binary as two stars at the same distance with the same age and metallicity, as expected if they formed together. The composite spectra are the sum of the two components,
\begin{align}
 F_{\lambda,\mathrm b}(M_1,q)&{}=F_{\lambda,1}+F_{\lambda,2}=F_\lambda(M_1)+F_\lambda(qM_1),\label{eq:sum}\\
 \frac{F_{\lambda,\mathrm b}}{F_{\lambda,1}}&{}=1+\frac{F_{\lambda,2}}{F_{\lambda,1}}.\label{eq:contrast}
\end{align}
The second equation gives the composite relative to the primary alone. A flux ratio $F_{\lambda,2}/F_{\lambda,1}$ that depends on wavelength changes the shape of the spectra, whereas two identical stars leave the shape unchanged and double the luminosity, a brightening of 0.753\,mag.

\begin{figure*}
\centering
\includegraphics[width=\textwidth]{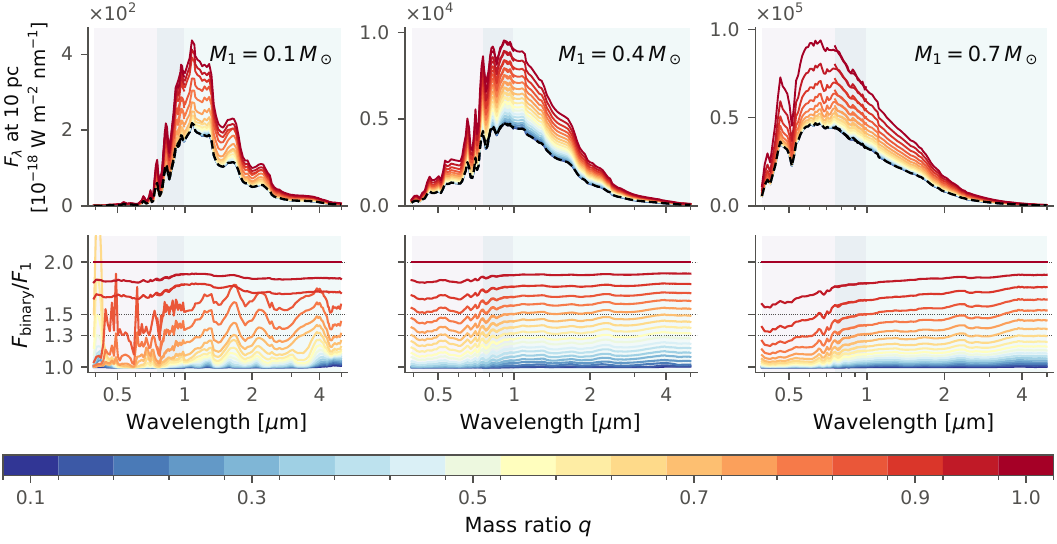}
\caption{Emulated binary spectra for primaries of 0.1, 0.4 and 0.7\,\Msun\ at 1\,Gyr and solar metallicity, for $q=0.10$--1.00 (colour). Top: composite flux at 10\,pc, with the primary alone dashed. Bottom: composite relative to the primary, $F_{\lambda,\mathrm b}/F_{\lambda,1}$ (Eq.~\ref{eq:contrast}; dotted: 1.3, 1.5 and 2). The shaded backgrounds mark the XP (purple) and \SPHEREx\ (teal) ranges, and where they overlap at 0.75--0.99$\mum$ the two are drawn as separate lines. The spike below 0.45$\mum$ at $M_1=0.1\,\Msun$ comes from the unvalidated blue flux of the theoretical models.}
\label{fig:binaryq}
\end{figure*}

Figure~\ref{fig:binaryq} shows both regimes. For $q\lesssim0.5$ the added flux rises from the blue end of XP into the \SPHEREx\ range, so much of the change in shape lies beyond 1$\mum$. Towards $q=1$ the added flux flattens, and the composite approaches twice the primary at all wavelengths. For the 0.1\,\Msun\ primary the companions come from the theoretical models, and their molecular bands imprint on the composite.

Both effects resemble a change of metallicity. At a fixed spectral shape a metal-rich MS star is more luminous than a metal-poor one, so a single-star model can absorb the extra light of a near-equal-mass companion with a higher \mh\ and a slightly higher mass. A companion at $0.5\lesssim q\lesssim0.8$ also makes the composite redder, which the single-star model again absorbs with a higher metallicity. Single-star fits to binaries therefore overestimate the metallicity by a few tenths of a dex, and near $q=1$ also the primary mass. The degeneracy is strongest for M dwarfs, where XPS cannot exclude a twin companion.

A companion may itself be an unresolved close pair, which makes the system a hierarchical triple. We model this case by adding a third coeval component to the modelled flux, $F_\lambda(M_1)+F_\lambda(M_{2a})+F_\lambda(M_{2b})$ at one age and metallicity.

\subsection{Validation}\label{sec:validation}

The emulator is validated on stars that were not used in training (Table~\ref{tab:training}). For the cool and warm validation stars the data-driven model reproduces the observed spectra to a median per-star residual below 4 per cent in every band, while the seven ultracool validation stars leave larger residuals, especially in XP. Fits of observed GCNS stars behave the same way: for typical single stars near the ridge at 0.5 and 0.3\,\Msun, the median absolute residual is 1.4 and 1.8 per cent in XP and 1.7 and 2.9 per cent in \SPHEREx. At 0.1\,\Msun\ the blue XP flux has a low S/N, and the model is about 10 per cent too bright at 0.9--1.3$\mum$. Where XP and \SPHEREx\ overlap, the two outputs of the network agree to a few per cent. The theoretical models are less accurate. For 13 field brown dwarfs at 1600--2000\,K, the median absolute fractional residuals in the \SPHEREx\ bands are 17--20 per cent for BT-Settl and 23--28 per cent for ATMO\,2020 at fixed literature labels (Appendix~\ref{app:physical}).

\begin{table}
\caption{Median per-star residual of the data-driven model on the validation stars.}\label{tab:training}
\centering
\small\setlength{\tabcolsep}{3pt}
\begin{tabular}{lrrrrr}
\toprule
Sample & $N$ & XP blue & XP red & 1--2$\mum$ & 2--5$\mum$ \\
\midrule
Cool (literature) & 213 & 2.6\% & 1.2\% & 3.3\% & 2.8\% \\
Warm (APOGEE) & 1462 & 1.0\% & 0.8\% & 3.7\% & 2.5\% \\
Ultracool & 7 & 14.4\% & 7.7\% & 3.2\% & 5.2\% \\
\bottomrule
\end{tabular}
\tablefoot{Median over stars of the per-star median absolute fractional residual in high-S/N channels.}
\end{table}

\subsection{Likelihood for observed spectra}\label{sec:inference}\label{sec:chi2}\label{sec:realroute}

We fit each star as a single star and as a binary, and measure how much the companion improves the fit. A single star is fitted over $(M,\mh,\tau)$ and a binary over $(M_1,q,\mh,\tau)$, with one age and metallicity for both components, on the absolute linear fluxes of XP alone or of XPS. Extinction is fixed at zero for the GCNS sample within 100\,pc, where the three-dimensional dust map of \citet{Edenhofer2024} shows little dust, whereas the upper-limit fits of the more distant \Gaia\ DR3 orbits and literature systems fit $A_V$ as well.

Observed spectra differ from the emulator by more than their measurement errors. The validation stars put this model error below 2 per cent per channel in XP and at 4--9 per cent in \SPHEREx. It contains the error of the emulator as well as the calibration and extraction of the \SPHEREx\ Quick Release spectra. The likelihood of a model therefore includes it,
\begin{equation}\label{eq:lnl}
 -2\ln\mathcal L=\boldsymbol\delta^{\mathsf T}\mathbf C^{-1}\boldsymbol\delta+\ln\det\mathbf C,\qquad \mathbf C=\mathbf C_{\rm meas}+\sigma_s^2\,\boldsymbol F\boldsymbol F^{\mathsf T}+\mathbf C_{\rm model}.
\end{equation}
Here $\boldsymbol\delta$ is the residual between the observed flux vector $\boldsymbol F$ and the model, both at 10\,pc, and $\mathbf C_{\rm meas}$ is the diagonal measurement covariance, which neglects the correlation between XP samples reconstructed from common coefficients \citep{Zhang2023}. The term $\sigma_s=2\sigma_\varpi/\varpi$ is the relative uncertainty of the flux scale that the parallax shares among all channels. The model covariance is fractional and of low rank,
\begin{equation}\label{eq:cmodel}
 \mathbf C_{\rm model}=\mathrm{diag}(\boldsymbol F_{\rm mod})\bigl(\mathbf B\mathbf B^{\mathsf T}+\mathrm{diag}\,\boldsymbol d^2\bigr)\,\mathrm{diag}(\boldsymbol F_{\rm mod}),
\end{equation}
where the columns of $\mathbf B$ are the leading eigenvectors of the covariance of $\ln(F_{\rm obs}/F_{\rm mod})$ of the validation stars, scaled by the square roots of their eigenvalues, and $\boldsymbol d$ is the standard deviation of the remaining scatter of each channel. The data-driven model and the theoretical models each have their own model covariance, evaluated for a binary on the summed flux of the components that each predicts, so that two components from the same model have fully correlated errors and components from different models independent ones.

Because the single-star and binary models have different model covariances, we keep the log determinant and measure the preference for a companion by
\begin{equation}\label{eq:dchi2}
 D=(-2\ln\mathcal L)_{\rm s}-(-2\ln\mathcal L)_{\rm b},
\end{equation}
the improvement in $-2\ln\mathcal L$ that a companion provides. We call a star detected when $D>25$, or $\dchi>25$ for mock spectra with measurement noise only, and undetected otherwise. None of the simulated single stars reaches this threshold (Appendix~\ref{app:mocks}), whereas about one in ten observed stars classified as single from RVs does (Sect.~\ref{sec:gcns}), which sets the false-positive rate (FPR) of a detection.

\label{sec:profiles}For the GCNS stars we also refit the binary at 19 fixed mass ratios from 0.1 to 1 and take the profile $\lambda(q)$, the excess of $-2\ln\mathcal L$ over the best model. A mass ratio is allowed when $\lambda(q)\le K(q)\,t$, where the threshold $t$ and the factor $K$ are calibrated on mock and observed binaries so that the true $q$ is excluded for 5--6 per cent of them (see more details in Appendix~\ref{app:mock}).

\subsection{Test on observed spectra}\label{sec:real}

We test the fits on the benchmark binaries of Sect.~\ref{sec:sample}, whose mass ratios are known independently. On the 816 benchmark binaries and their magnitude- and colour-matched field controls, XP alone provides most of the separation. Adding \SPHEREx\ raises the median $D$ of the binaries by a factor of 1.5--2.2, but also raises the fraction of controls above $D=25$ from 16 to 25 per cent. The probability that a binary ranks above a control of the same catalogue (the area under the receiver operating characteristic curve, AUC; \citealt{Hanley1982,Fawcett2006}) does not improve.

\begin{table}
\caption{Benchmark sample and results of the paired fits with XP alone.}\label{tab:bench}
\centering
\small\setlength{\tabcolsep}{2.2pt}
\begin{tabular}{lrrcrr}
\toprule
Catalogue & $N_{\rm b}$ & $N_{\rm c}$ & $q$ & AUC & med.\ $|\Delta q|$ \\
\midrule
Gaia SB1 & 742 & 1484 & 0.07--0.99 & 0.78 & 0.15 \\
Gaia SB2 & 35 & 70 & 0.73--1.00 & 0.94 & 0.09 \\
APOGEE SB2 & 30 & 60 & 0.57--0.99 & 0.93 & 0.07 \\
DEBCat & 8 & 16 & 0.44--1.00 & 0.97 & 0.09 \\
Gaia eclipsing binary & 1 & 2 & -- & -- & -- \\
\bottomrule
\end{tabular}
\tablefoot{$N_{\rm b}$, $N_{\rm c}$: binaries and matched field controls. AUC: probability that a binary has a larger $D$ than a control of the same catalogue. $|\Delta q|$: absolute difference between fitted and external mass ratio. Grid fits.  Adding \SPHEREx\ raises the median $D$ of the binaries by a factor of 1.5 for APOGEE SB2, 2.2 for Gaia SB2 and 1.6 for Gaia SB1. With \SPHEREx\ the AUC is 0.76, 0.93, 0.91 and 0.82 and the median $|\Delta q|$ 0.19, 0.09, 0.09 and 0.19 for Gaia SB1, Gaia SB2, APOGEE SB2 and DEBCat.}
\end{table}

\label{sec:pilot}With XP alone this probability is 0.93--0.97 for SB2 and eclipsing binaries and 0.78 for \Gaia\ SB1 systems, whose companions are generally fainter (Table~\ref{tab:bench}). Spectral mass ratios cluster towards 0.6--0.8: faint companions change the spectra little, and a single star of higher metallicity can absorb part of the light of a near-equal pair.

\begin{figure*}
\centering
\includegraphics[width=\textwidth]{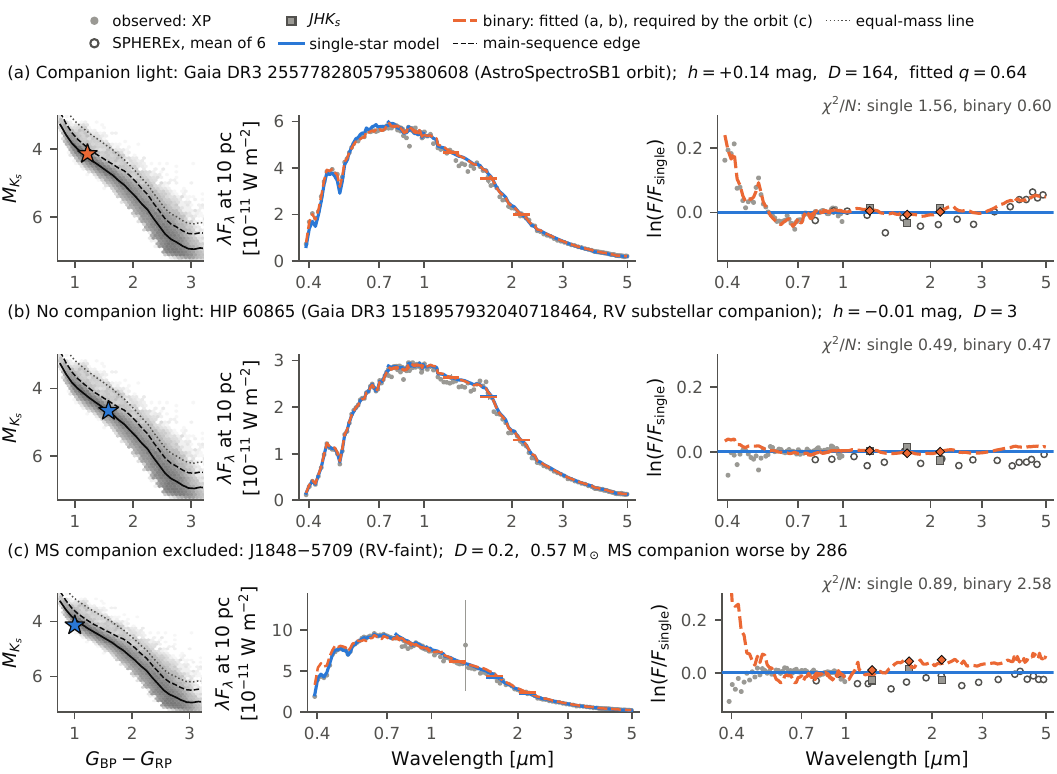}
\caption{Fits to three stars with RV follow-up. (a) Companion light ($D=164$; Eq.~\eqref{eq:dchi2}). In each of three colour bins we selected the independently known binary within 0.15\,mag of the ridge with $D$ closest to the bin median. The star shown is from the bluest bin. (b) No companion light: HIP~60865 hosts a substellar companion \citep{Barbato2026}. (c) MS companion excluded: for J1848$-$5709 \citep{Yamaguchi2024} the orange model is refitted with an MS companion that satisfies the mass function of the \Gaia\ orbit (0.57\,\Msun). Left: CMD with the ridge (solid), upper edge of the MS (dashed) and equal-mass line (dotted). Middle: flux at 10\,pc (grey), single-star model (blue) and binary model (orange). Right: data and binary model relative to the single-star model, on a common scale. Open circles average six \SPHEREx\ channels, squares mark $JHK_s$ and diamonds the binary model in $JHK_s$.}
\label{fig:outcomes}
\end{figure*}

XPS detects companion light in the binary on the MS ridge (Fig.~\ref{fig:outcomes}a). For HIP~60865 the single star fits the spectra despite its known substellar companion (b). For J1848$-$5709 an MS companion that satisfies the mass function of the \Gaia\ orbit would add about 5 per cent to the flux in $H$ and $K_s$, while the spectra exclude it (c).

\section{Results}\label{sec:results}

The fits of Sect.~\ref{sec:model} give two quantities for each star: $D$, which measures the preference for a binary over a single star, and the profile $\lambda(q)$, which shows the mass ratios that the spectra allow. We apply $D$ to the GCNS to find binaries that the CMD misses (Sect.~\ref{sec:gcns}), and compare the fits at the two mass ratios that a photocentre orbit allows (Sect.~\ref{sec:gcns-orbits}). Where $D$ shows no companion, $\lambda(q)$ limits its light, and together with the orbital mass this limit identifies companions too faint to be MS stars (Sect.~\ref{sec:gcns-limits}).

\subsection{Binaries that the CMD misses}\label{sec:gcns}

Binaries are usually selected in the CMD: stars above the single-star sequence are counted as binaries, either with a cut in height or with a mixture model of single stars and binaries \citep[as in open clusters;][]{LiShao2022}. In the field the sequence is broadened by metallicity, and a binary can hide in it for two reasons: its companion adds too little light to move it, or the companion lifts a metal-poor primary onto the sequence of more metal-rich single stars. We test whether fits to the absolute spectra find the binaries that a cut on CMD height misses.

\begin{figure*}
\centering
\includegraphics[width=\textwidth]{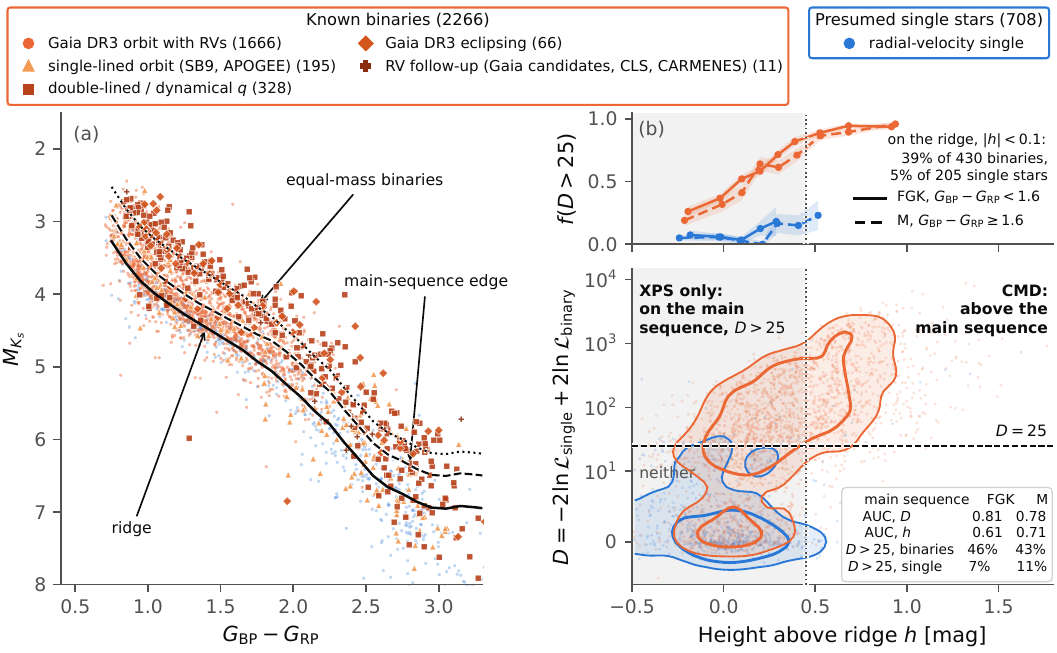}
\caption{Known binaries (orange) and presumed single stars (blue) from the literature, at $9\le G<17.7$. (a) CMD with the ridge (solid; median \MKs\ of the stars of the GCNS sample without a \Gaia\ non-single-star solution, so that the plotted binaries lie mostly above it), the upper edge of the MS (dashed) and the equal-mass line (dotted). (b) $D$ against height $h$ above the ridge (bottom; contours enclose 50 and 80 per cent, the shaded band is the widest extent of the MS) and the fraction with $D>25$ (top). The box gives the AUC of $D$ and $h$ on the MS for FGK ($BP-RP<1.6$) and M stars.}
\label{fig:literature}
\end{figure*}

We define the MS band up to $\min(2\sigma,0.45\,\mathrm{mag})$ above the ridge, using the robust scatter $\sigma$ of stars without a \Gaia\ non-single-star solution in 0.3\,mag bins of $BP-RP$. Of the literature binaries, 58 per cent lie within this band (Fig.~\ref{fig:literature}a). A cut on height at 0.29\,mag has the same 11 per cent false-positive rate on RV single stars as $D>25$. $D>25$ recovers 44 per cent of the binaries missed by this cut and 86 per cent of those it selects (Table~\ref{tab:literature} and Fig.~\ref{fig:literature}b).

\begin{table}
\caption{Literature binaries selected by $D>25$ and by height above the ridge.}\label{tab:literature}
\centering
\small\setlength{\tabcolsep}{4pt}
\begin{tabular}{lrrr}
\toprule
Sample & FPR & $D>25$ & height \\
\midrule
\multicolumn{4}{l}{\emph{$9\le G<17.7$}} \\
All & 10.7 & 64 & 48 \\
FGK & 8.1 & 64 & 48 \\
M & 12 & 64 & 55 \\
\midrule
\multicolumn{4}{l}{\emph{All $G$}} \\
All & 6.5 & 59 & 20 \\
FGK & 5.2 & 57 & 14 \\
M & 11 & 64 & 55 \\
\midrule
\multicolumn{4}{l}{\emph{Dynamical mass ratio, all $G$}} \\
$0.2\le q<0.5$ & & 41 & 24 \\
$0.5\le q<0.8$ & & 69 & 50 \\
$0.8\le q<0.95$ & & 76 & 64 \\
$q\ge0.95$ & & 87 & 68 \\
\midrule
\multicolumn{4}{l}{\emph{AUC on the MS, all $G$}} \\
FGK & & 0.80 & 0.51 \\
M & & 0.79 & 0.70 \\
\bottomrule
\end{tabular}
\tablefoot{Per cent, except the AUC. FPR: fraction selected among stars classified as single from RVs, equal for both scores. The height cut that gives this FPR is 0.29, 0.25 and 0.30\,mag for all, FGK and M stars at $9\le G<17.7$, and 0.68, 0.76 and 0.31\,mag at all $G$. The rows by mass ratio use the FPR of each colour range.}
\end{table}

\begin{figure*}
\centering
\includegraphics[width=\textwidth]{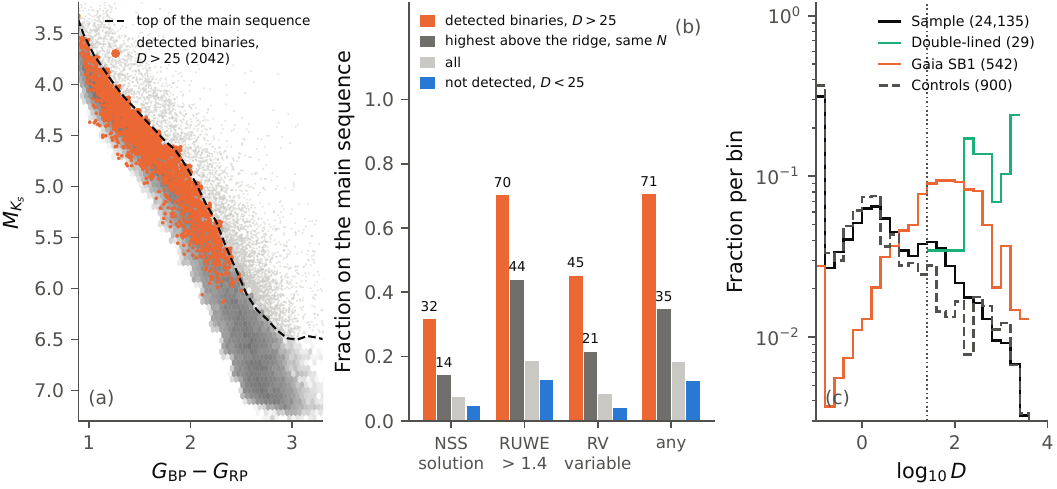}
\caption{Binary candidates in the GCNS at 0.5--0.9\,\Msun. (a) CMD; orange: detected binaries ($D>25$) on the MS (below the dashed line). (b) Fraction of MS stars with a binary signature independent of XPS, for the detected binaries ($D>25$), the same number of stars highest above the ridge, all MS stars and the stars without a detection ($D<25$). (c) Distribution of $D$ for the sample, the SB2 (green) and \Gaia\ SB1 (orange) benchmark binaries and their controls (dashed); dotted: $D=25$.}
\label{fig:gcns-candidates}
\end{figure*}

Applied to the GCNS stars of 0.5--0.9\,\Msun, the same selection gives 2042 candidates with $D>25$ within the MS band (Fig.~\ref{fig:gcns-candidates}a). Of the candidates, 71 per cent show at least one sign of binarity that uses neither XPS nor the CMD, such as a \Gaia\ non-single-star solution, a renormalised unit weight error $\mathrm{RUWE}>1.4$ or variable \Gaia\ RVs, twice the 35 per cent of the same number of MS stars highest above the ridge (Fig.~\ref{fig:gcns-candidates}b). Most benchmark controls that pass $D>25$ also carry these signs and are probably unrecognised binaries (Fig.~\ref{fig:gcns-candidates}c).

\begin{figure*}
\centering
\includegraphics[width=\textwidth]{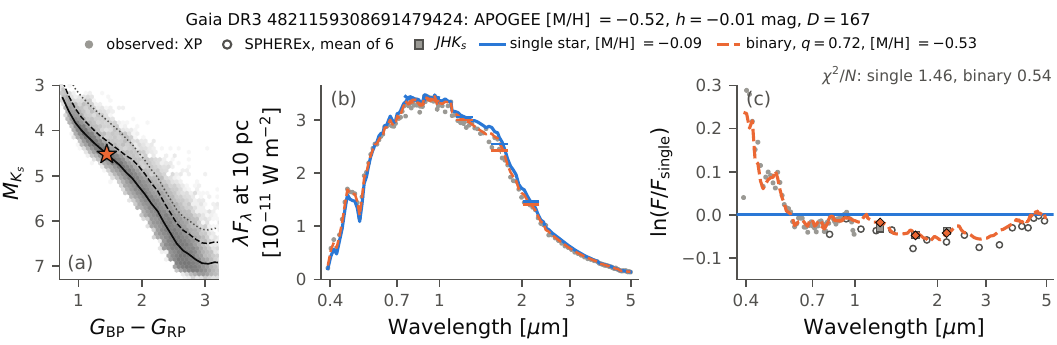}
\caption{A metal-poor binary on the MS ridge, \Gaia\ DR3 4821159308691479424, which has an \texttt{AstroSpectroSB1} orbit and an APOGEE metallicity of $-0.52$. (a) CMD position, with the ridge (solid), the upper edge of the MS (dashed) and the equal-mass line (dotted). (b) $\lambda F_\lambda$ at 10\,pc (grey) with the best single-star model (blue, $\mh=-0.09$, $\chi^2/N=1.46$) and the best binary model (orange, $q=0.72$, $\mh=-0.53$, $\chi^2/N=0.54$). (c) Data and binary model relative to the single-star model, which falls short of the XP flux below 0.55$\mum$ by up to a third. Open circles average six \SPHEREx\ channels, diamonds mark the binary model in $JHK_s$.}
\label{fig:metallicity-example}
\end{figure*}

XPS also shows why these binaries hide. The candidates with an APOGEE metallicity have a median $\mh=-0.22$, against $-0.08$ for the stars with $D<25$, so their primaries are metal-poor and would lie below the ridge. Relative to stars of the same APOGEE metallicity they are 0.18\,mag too bright, close to the 0.21\,mag that the fitted companions add. In the CMD such a pair looks like a single star of higher metallicity, and a single-star fit to the spectra makes the same mistake: it returns a metallicity 0.31\,dex above the APOGEE value. The binary fit agrees with APOGEE to within 0.11\,dex. Figure~\ref{fig:metallicity-example} shows one such system.

This result holds for primaries of 0.5--0.9\,\Msun. At 0.3--0.5\,\Msun, 20 per cent of the candidates within the MS band show an independent sign of binarity, no more than the 25 per cent of the same number of stars highest above the ridge. This is also the mass range where a twin companion is most easily mistaken for a higher metallicity (Sect.~\ref{sec:binary}). Appendix~\ref{sec:lit-details} describes the literature samples.

\subsection{Mass ratios from \Gaia\ photocentre orbits}\label{sec:gcns-orbits}

A small \Gaia\ photocentre orbit can arise from either a low-mass dark companion or a luminous companion of nearly equal mass. Distinguishing these possibilities requires knowing how much light the companion contributes. \Gaia\ follows the photocentre of the pair. When the two stars have similar masses and luminosities, the photocentre lies close to the centre of mass and moves little even though the companion is massive. The astrometric mass-ratio function (AMRF; \citealt{Shahaf2019}) relates the orbit to the mass ratio and flux ratio,
\begin{equation}\label{eq:amrf}
 \mathcal A\equiv\frac{a_0}{\varpi}\left(\frac{M_1}{\Msun}\right)^{-1/3}\left(\frac{P}{\rm yr}\right)^{-2/3}=\frac{q-\beta}{(1+q)^{2/3}(1+\beta)},
\end{equation}
where $a_0$ is the angular semi-major axis of the photocentre, $\varpi$ the parallax, $M_1$ the primary mass, $P$ the period, $q=M_2/M_1$ the mass ratio and $\beta=F_2/F_1$ the flux ratio in the \Gaia\ $G$ band. Given the primary mass, the orbit constrains a combination of $q$ and $\beta$, so it cannot determine the mass ratio alone \citep{Halbwachs2023}. Assuming a dark companion gives the minimum mass ratio $q_{\rm dark}$. The more light the companion contributes, the larger the mass ratio required by the same orbit. Ignoring this light can therefore make a near-equal-mass binary appear to host a low-mass companion. \citet{BailerJones2026} distinguish these possibilities with the mass--luminosity relation of an isochrone and \Gaia\ photometry. Here we use XPS spectra to constrain the companion light and determine which orbital solution agrees with the observations.

\begin{figure*}
\centering
\includegraphics[width=.28\textwidth]{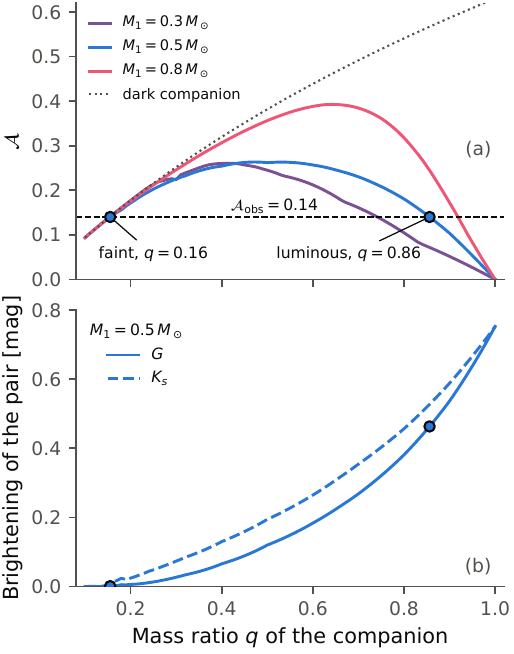}\hfill\includegraphics[width=.70\textwidth]{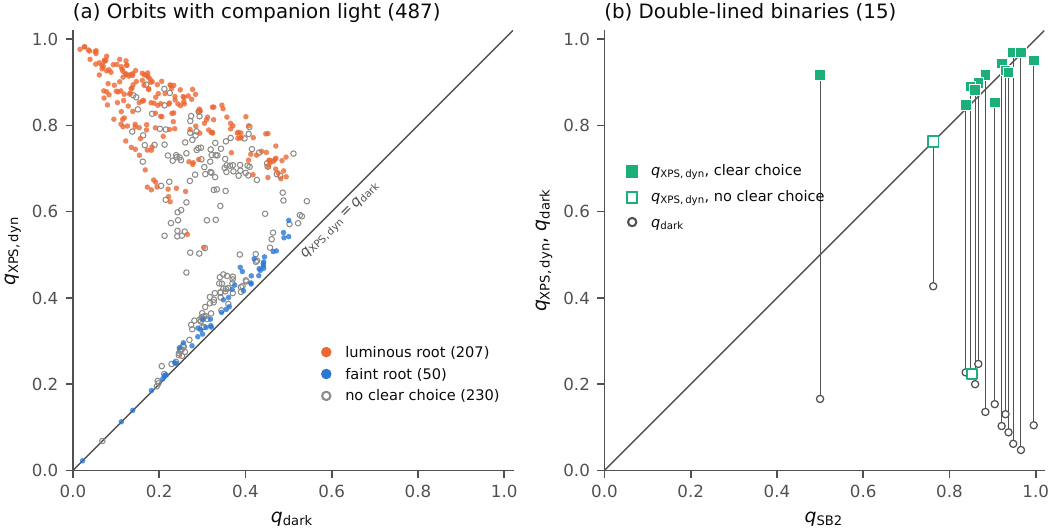}
\caption{Photocentre orbits and companion light. Left: (a) the same orbit ($\mathcal A=0.14$, dashed) admits a faint and a luminous companion to a $0.5\,\Msun$ primary. The dotted curve assumes a dark companion. (b) Brightening in $G$ and $K_s$. Right: (a) the mass ratio at the preferred root against the dark-companion estimate for 487 GCNS orbits with $D>25$. Filled symbols mark a difference above 25 in $-2\ln\mathcal L$ (orange: luminous root; blue: faint root). (b) Comparison with 15 SB2 mass ratios. Squares use the root preferred by XPS (open for an ambiguous choice). Circles assume a dark companion.}
\label{fig:orbit-roots}
\end{figure*}

XPS can distinguish a faint companion from a luminous, near-equal-mass companion by comparing the spectra predicted at the two orbital roots. For the primary of the free binary fit, the emulator gives the $G$-band flux ratio $\beta_G(q)$ of a coeval companion. As $q$ increases, $\mathcal A$ first rises. When the luminosities of the two stars become similar, the photocentre motion diminishes and $\mathcal A$ tends back towards zero. An observed $\mathcal A$ below the maximum of this curve, $\mathcal A_{\rm max}$, therefore generally admits two mass ratios: a faint root near $q_{\rm dark}$ and a luminous root of nearly equal mass (Fig.~\ref{fig:orbit-roots}, left). We refit XPS with $q$ fixed at each root and denote the better-fitting root by $q_{\rm XPS,dyn}$ (Table~\ref{tab:symbols}). The difference in $-2\ln\mathcal L$ measures the spectral preference for one root over the other.

The APOGEE SB2 system \Gaia\ DR3 1303724304656566528 illustrates this test. Its two orbital roots lie at $q=0.16$ and 0.85. The faint root leaves negative residuals in the blue XP samples. These residuals vanish at the luminous root, which lowers $-2\ln\mathcal L$ by 121. The spectrally preferred root is close to the independent APOGEE measurement of $q=0.905$ \citep{Kounkel2021}. For this system the faint root would assign a low mass to a companion nearly as massive as its primary.

\begin{table}
\caption{Companion mass ratios and flux ratios used in the orbit analysis.}\label{tab:symbols}
\centering
\small\setlength{\tabcolsep}{3pt}
\begin{tabular}{@{}lp{6.1cm}@{}}
\toprule
Symbol & Definition \\
\midrule
$q_{\rm dark}$ & Mass ratio inferred from the photocentre orbit assuming a dark companion. \\
$q_{\rm XPS}$ & Mass ratio fitted to the spectra alone. \\
$q_{\rm XPS,dyn}$ & Mass ratio at the orbital root that best fits the spectra. \\
$q_{\rm dyn}$, $\beta_{\rm dyn}$ & Mass ratio and $G$-band flux ratio inferred from the RV and photocentre orbits for the adopted primary mass. \\
\midrule
$\beta_G(q)$ & Model $G$-band flux ratio ($F_2/F_1$) for a coeval companion at mass ratio $q$. \\
$q_{\rm up}$, $\beta_{\rm up}$ & Upper limits on the mass ratio and $G$-band flux ratio of a coeval companion from the spectral fits. \\
$M_{2,\rm dark}$ & Companion mass inferred from the orbit assuming zero companion light. \\
$M_{2,\rm up}$ & Largest mass of a single MS companion allowed by the spectra. \\
$\beta_{\rm MS}$ & Predicted $G$-band flux ratio for an MS companion of the orbital mass. \\
\bottomrule
\end{tabular}
\end{table}

SB2 systems test whether XPS chooses the right orbital root. The velocities of both stars give their mass ratio without assuming how much light the companion contributes. The parent sample contains 15 SB2 systems with a \Gaia\ astrometric orbit. All but one have $q\ge0.76$, but assuming a dark companion generally gives much smaller mass ratios. The median difference is 0.75 (Fig.~\ref{fig:orbit-roots}, right).

XPS clearly chooses the luminous root for 13 of the 15 systems. For 12 of these 13, $q_{\rm XPS,dyn}$ agrees with the SB2 value to within 0.055. Adding the orbit also improves the mass ratio: for most of the \Gaia\ systems with a clear choice, $q_{\rm XPS,dyn}$ is closer to the SB2 value than the spectra-only estimate $q_{\rm XPS}$. At the SB2 mass ratios $q=0.84$--1, however, the model $G$-band flux ratio remains about 12 per cent below the value required by the photocentre orbit.

Next, we test intermediate mass ratios with the 953 \texttt{AstroSpectroSB1} systems of the parent sample. These systems have both RV and astrometric orbits. The RV orbit gives the projected size of the primary orbit, $a_1\sin i$, and the astrometry gives the inclination $i$ and the photocentre orbit size $a_0$. Together with the primary mass from the free binary fit, these measurements give the mass ratio $q_{\rm dyn}$ and the companion-to-primary flux ratio $\beta_{\rm dyn}$ required to explain the photocentre motion.

At $q_{\rm dyn}=0.3$--0.7, the flux ratios inferred from the orbits agree with those predicted by the emulator. Above $q_{\rm dyn}=0.7$, however, the two differ substantially: the orbit gives $\beta_{\rm dyn}=0.02$, against a model prediction of 0.18.

Of the 487 GCNS orbits with detected companion light, 257 have a difference greater than 25 in $-2\ln\mathcal L$ between the two roots. Of these, 207 favour the luminous root. Their median mass ratio is 0.19 under the dark-companion assumption and 0.84 at the luminous root. Ignoring the companion light would therefore assign low masses to many companions nearly as massive as their primaries.

\subsubsection{Luminous companions among substellar candidates}\label{sec:orbit-tool}

Searches for brown dwarfs and giant planets in \Gaia\ orbits usually select systems with no obvious sign of a companion and small photocentre orbits. RV follow-up has nevertheless shown many of these candidates to be near-equal-mass binaries \citep{MarcussenAlbrecht2023,Unger2023,Barbato2026,Marcussen2026}.

\begin{figure}
\centering
\includegraphics[width=\columnwidth]{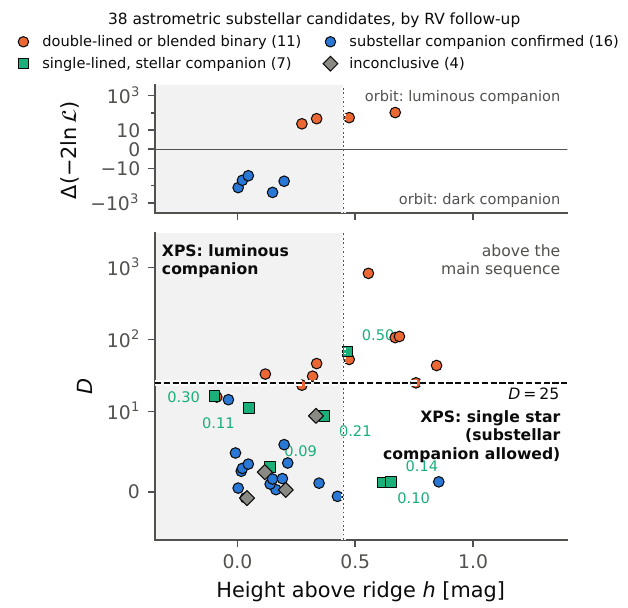}
\caption{\Gaia\ DR3 astrometric substellar candidates with RV follow-up. Bottom: $D$ against height above the ridge for the 38 with an XPS fit, SB1 systems labelled with $q$. The 11 binaries identified from double or asymmetric lines have $D=16$--826, the 16 stars with a confirmed substellar companion $D\le15$. The dashed line marks $D=25$, the dotted line the upper edge of the MS. Top: $-2\ln\mathcal L$ of the dark minus the luminous branch (Sect.~\ref{sec:gcns-orbits}) for the nine targets with an orbit fit (four binaries, five substellar hosts). Positive values prefer a luminous companion.}
\label{fig:bd-candidates}
\end{figure}

We fit the XPS spectra of 38 of these systems (Fig.~\ref{fig:bd-candidates}). RV observations identify 11 binaries through double or asymmetric lines, with $D=16$--826. Another 16 systems host confirmed substellar companions, all with $D\le15$. All of these binaries therefore have larger $D$ than every substellar host, while the two groups overlap in height above the CMD ridge. We also fit the orbits of nine of these systems and use XPS to select the correct root in all nine: the binaries favour the luminous root and the substellar hosts favour the faint root. We can thus separate the two groups in this sample by ranking $D$, although applying the detection threshold $D>25$ would miss the binary with the weakest signal.

If we ignore companion light, we can also mistake a broad range of companion masses for a concentration near the hydrogen-burning limit. Near $\mathcal A_{\rm max}$ the curve is flat, so luminous companions of different masses give similar $\mathcal A$. Assuming that these companions are dark then gives similar inferred masses. For a primary of 0.35\,\Msun, companions of 0.08--0.21\,\Msun\ all give $\mathcal A$ close to $\mathcal A_{\rm max}$. Under the dark-companion assumption, this maximum corresponds to a companion mass of 0.10\,\Msun.

We find that the orbits pile up near their respective $\mathcal A_{\rm max}$ at different primary masses. The corresponding dark-companion mass is not fixed: as the primary mass increases from about 0.3 to 1.1\,\Msun, it rises from 0.09 to 0.57\,\Msun. Only for M-dwarf primaries does this location approach the hydrogen-burning limit. We therefore cannot infer a population of brown dwarfs from a concentration of inferred masses there alone. For FGK primaries the situation differs. Both M dwarfs near the hydrogen-burning limit and brown dwarfs are much fainter than the primary, with $\beta\ll q$, so ignoring their light still gives masses close to the true values.

We can use XPS to remove luminous companions from the substellar candidates. Of the 879 systems with $M_{2,\rm dark}<0.08$\,\Msun, we detect companion light in 224. These include all six substellar candidates that RV observations identify as SB2 or blended binaries. For the substellar companion systems confirmed by RV follow-up in the preceding comparison, we detect no companion light. The orbits of the remaining candidates, however, allow either a brown dwarf or an M dwarf of 0.09--0.14\,\Msun. We cannot currently distinguish these possibilities with XPS.

\subsection{Upper limits on luminous companions}\label{sec:gcns-limits}

Even when we detect no companion light, we can constrain its brightness and combine this limit with the orbit to test whether the companion can be an MS star. We derive these limits and apply them to nearby orbits (Sect.~\ref{sec:too-faint}), extend them to all \Gaia\ DR3 astrometric orbits (Sect.~\ref{sec:nss-catalogue}) and compare the selection with the colour-excess test (Sect.~\ref{sec:colour-test}).

\subsubsection{Companions too faint for their mass}\label{sec:too-faint}

We use the mass-ratio profile of Sect.~\ref{sec:profiles} to find the mass ratios that satisfy the calibrated threshold $\lambda(q)\le K(q)\,t$. We define the flux-ratio and mass-ratio upper limits as
\begin{equation}\label{eq:betaup}
 \beta_{\rm up}=\max_{\lambda(q)\le K(q)\,t}\beta_G(q),
 \qquad
 q_{\rm up}=\max_{\lambda(q)\le K(q)\,t}q.
\end{equation}
Here $\beta_G(q)$ is the companion-to-primary $G$-band flux ratio that we obtain after refitting the primary at fixed $q$. We interpolate to locate the boundaries of the allowed ranges between adjacent profile nodes and include these boundaries when calculating both upper limits.

If the system also has an orbit, we can combine it with the flux-ratio upper limit to constrain the mass ratio without first choosing the faint or luminous root,
\[
 q_{\rm dark}\le q\le q(\mathcal A_{\rm obs},\beta_{\rm up}).
\]
The lower bound corresponds to a dark companion and the upper bound to the largest flux ratio that the spectra allow. If every MS root of the orbit requires more companion light than $\beta_{\rm up}$ allows, we can exclude a single MS companion. We can also distinguish a preference for one root from exclusion of the other. Of the 207 systems that favour the luminous root in Sect.~\ref{sec:gcns-orbits}, we exclude the low-mass, faint root in 96 using the full calibrated mass-ratio profile.

For undetected primaries of 0.7--0.9\,\Msun, the median $q_{\rm up}$ is 0.31 and the median $\beta_{\rm up}$ is 0.008. As an independent check, none of the 175 undetected \texttt{AstroSpectroSB1} systems in the calibrated primary-mass bins has a dynamical flux ratio more than twice its measurement error above $\beta_{\rm up}$. Applying the limits to the GCNS orbits, we exclude the luminous root in 78 systems and both roots in another four. We also exclude a single MS companion at the orbital mass for all undetected \texttt{AstroSpectroSB1} systems with $M_2\ge0.45$\,\Msun. Table~\ref{tab:limits} lists the outcomes.

\begin{figure*}
\centering
\includegraphics[width=\textwidth]{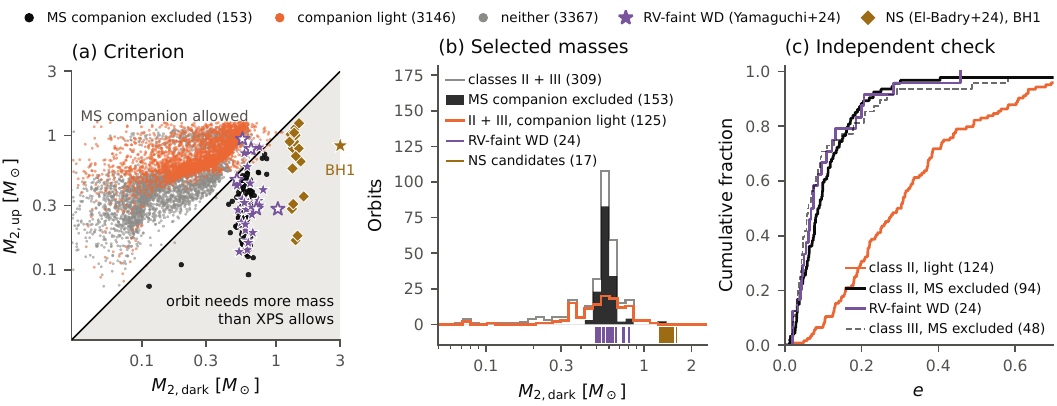}
\caption{Companions too faint for their mass among the \Gaia\ DR3 astrometric orbits within 200\,pc. (a) The criterion: the dark-companion mass $M_{2,\rm dark}$, the lowest mass that the orbit allows, against $M_{2,\rm up}$, the largest MS companion mass that XPS allows. In the shaded region an MS companion of the orbital mass is brighter than XPS allows. Black: $M_{2,\rm dark}$ exceeds $M_{2,\rm up}$ by more than twice its error (MS companion excluded); orange: companion light ($D>25$); grey: neither. Literature systems at their published companion mass: RV-faint companions (purple stars; filled: Group~1, open: Group~2; \citealt{Yamaguchi2024}), neutron-star candidates (ochre diamonds; \citealt{ElBadry2024NS}) and Gaia BH1 (at the right edge). (b) Dark-companion masses of classes II and III of \citet{Shahaf2023} (grey), of the orbits with the MS companion excluded (black) and of the class-II and class-III orbits with companion light (orange); ticks: the published masses of the RV-faint companions and the neutron-star candidates. (c) Cumulative distribution of the orbital eccentricity $e$ for class~II by outcome, for the RV-faint companions and for the excluded class-III orbits (dashed).}
\label{fig:dark-companions}
\end{figure*}

For the 6666 \Gaia\ DR3 orbits within 200\,pc, we refit the primary at a series of fixed companion masses and denote the largest MS companion mass allowed at $K(q)\,t$ by $M_{2,\rm up}$. We call a system MS-excluded when its dark-companion mass $M_{2,\rm dark}$, the minimum mass that the orbit allows, exceeds $M_{2,\rm up}$ by more than twice its catalogue mass error. We find 153 such systems (Fig.~\ref{fig:dark-companions}a). Their companion masses cluster at 0.53--0.62\,\Msun, the masses of typical white dwarfs. The lower edge of this range is partly set by our sensitivity. Among the 3514 orbits without detected companion light, the spectra would exclude an MS companion of 0.4\,\Msun\ in 35 per cent and one of 0.6\,\Msun\ in 78 per cent.

We test this criterion on compact-companion systems from the literature, including the neutron-star candidates of \citet{ElBadry2024NS} and Gaia BH1, BH2 and BH3 \citep{ElBadry2023BH1,ElBadry2023BH2,GaiaBH3}. We exclude an MS companion for all 17 neutron-star candidates with a usable limit and for Gaia BH1 (Table~\ref{tab:limits}).

The triage of \citet{Shahaf2023} uses an MS mass--luminosity relation to define three classes: a single MS companion can explain class~I, only an unresolved pair of MS stars can explain class~II, and even an MS pair cannot explain class~III. We exclude a single MS companion in 48 of the 50 class-III systems and in a further 11 class-I systems. Among the 259 class-II systems, whose orbits alone cannot distinguish a white dwarf from an unresolved stellar pair, we detect companion light in 124 and exclude a single MS companion in 94.

At the adopted primary mass, the RV and photocentre orbits of eight class-II or class-III systems independently require companion light, with $\beta_{\rm dyn}>2\sigma$. We detect companion light in all eight, with $D>25$. None of the 19 MS-excluded systems with the same orbital measurements shows a significant dynamical companion-light signal.

For the MS-excluded class-II systems, companion masses and orbital eccentricities together support a white-dwarf interpretation. Their companion masses are 0.54--0.64\,\Msun, close to the 0.54--0.69\,\Msun\ of the RV-faint white dwarfs. The two class-II groups have similar median periods ($\log_{10}(P/{\rm d})=2.69$ for MS exclusions and 2.66 for detections). The MS-excluded systems have a median eccentricity of 0.09 and 60 per cent have $e<0.1$. The class-II systems with companion light instead have a median eccentricity of 0.30 and only 7 per cent have $e<0.1$. The former group also resembles the MS-excluded class-III systems and the RV-faint white dwarfs, which also have low eccentricities (Fig.~\ref{fig:dark-companions}c). We do not use eccentricity in the selection, so these nearly circular orbits provide independent evidence consistent with circularisation during mass transfer as the white dwarf formed.

For the class-II systems with companion light, we instead find evidence for the mass of an unresolved stellar pair. Using the primary mass and flux ratio from XPS, we calculate the companion mass required by the orbit and compare it with the mass from a single-companion spectral fit. The median ratio is 1.53 for 121 detected class-II systems that allow a single MS companion but exclude a twin, compared with 0.99 for 2704 detected class-I systems. The light of the class-II companions corresponds to a less massive star than the orbit requires. An unresolved inner pair can supply both the mass and the light.

\begin{figure*}
\centering
\includegraphics[width=\textwidth]{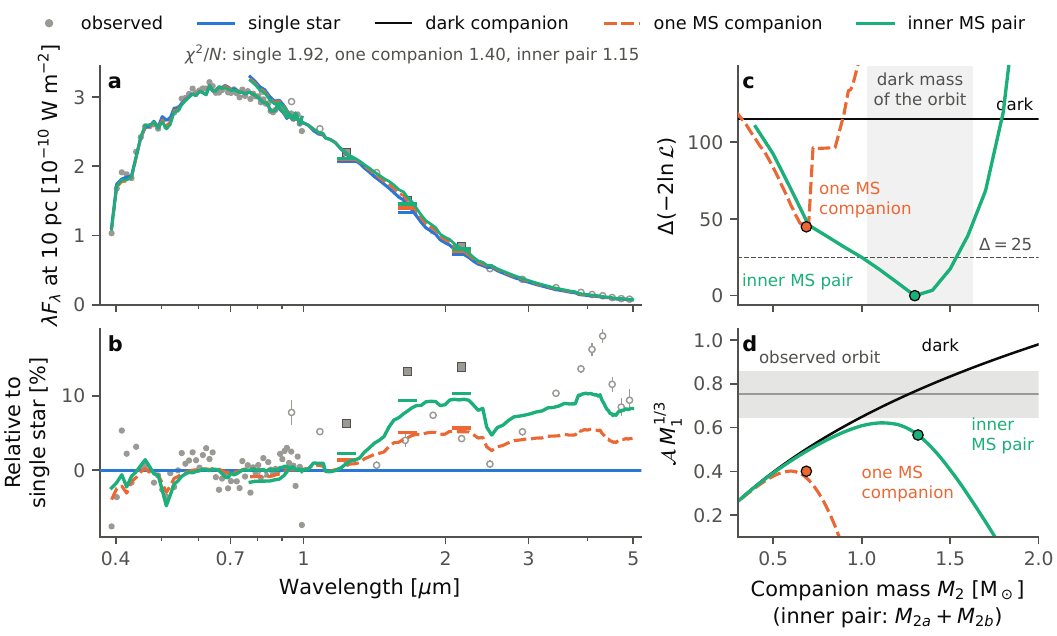}
\caption{The hierarchical triple \Gaia\ DR3 1748901959855337472 \citep{ElBadry2026} fitted as a single star (blue), with one MS companion (orange) and with an inner pair of MS stars (green). Both companion models include the photocentre orbit. (a) $\lambda F_\lambda$ at 10\,pc (filled: XP; open: \SPHEREx, means of six channels; squares: 2MASS). (b) The observed flux and the companion models relative to the single star. (c) $\Delta(-2\ln\mathcal L)$ against the companion mass, the total mass of the pair for the triple. Points mark the best fits, the horizontal line the dark solution, which the single star and the orbit satisfy exactly, and the shaded band the dark-companion mass of the orbit and its error. (d) The photocentre observable $\mathcal A\,M_1^{1/3}$ against the companion mass for a dark companion, one MS companion and an equal-mass MS pair, at the primary of the triple fit. The grey band is the observed orbit and the points the two fits. The eclipses have a period of 1.02\,d, and the pair model has two stars of 0.66\,\Msun.}
\label{fig:triple}
\end{figure*}

We test this interpretation directly by fitting the spectra and photocentre orbit together (Appendix~\ref{sec:inner-pairs}). The class-III system in Fig.~\ref{fig:triple} has eclipses that independently identify its companion as an inner binary \citep{ElBadry2026}. We sum the masses and spectra of the two inner stars and compare them with the observations. The pair gives a $G$-band flux ratio of 0.19, close to the $0.168\pm0.010$ inferred from the RVs and astrometry. Relative to one MS companion, the inner-pair model lowers $-2\ln\mathcal L$ by 45 in the joint fit.

\subsubsection{All \Gaia\ DR3 astrometric orbits}\label{sec:nss-catalogue}

\begin{table}
\caption{Outcome of the companion limits.}\label{tab:limits}
\centering
\footnotesize\setlength{\tabcolsep}{2pt}
\begin{tabular}{lrr}
\toprule
Sample & $N$ & Excluded \\
\midrule
\multicolumn{3}{l}{\emph{GCNS orbits with $D<25$, $G\ge9$, $0.3\le M_1\le0.9$\,\Msun}} \\
\texttt{Orbital}, no MS root$^{c}$ & 558 & 17 \\
\texttt{Orbital}, luminous root$^{c}$ & 558 & 78 \\
\texttt{Orbital}, both roots$^{c}$ & 558 & 4 \\
\texttt{AstroSpectroSB1}, $M_2\ge0.45$\,\Msun & 15 & 15 \\
\midrule
\multicolumn{3}{l}{\emph{Orbits within 200\,pc, by triage class}} \\
Class I & 6357 & 11 \\
Class II & 259 & 94 \\
Class III & 50 & 48 \\
\midrule
\multicolumn{3}{l}{\emph{All DR3 orbits of the triage, by class}} \\
Class I & 54\,402 & 272 \\
Class II & 9460 & 2097 \\
Class III & 829 & 472 \\
\midrule
\multicolumn{3}{l}{\emph{Literature systems with a limit}} \\
RV-faint companions, Group 1$^{a}$ & 24 & 22 \\
RV-faint, Group 2 & 5 & 3 \\
Neutron-star candidates$^{b}$ & 17 & 17 \\
Gaia BH1 & 1 & 1 \\
\bottomrule
\end{tabular}
\tablefoot{Upper block: Sect.~\ref{sec:too-faint}, orbits within 100\,pc. Second and last blocks: Fig.~\ref{fig:dark-companions}, where an excluded orbit is drawn black. The 153 MS-excluded orbits are the sum of the second block. Third block: Sect.~\ref{sec:nss-catalogue}. $^{a}$ \citet{Yamaguchi2024}. $^{b}$ \citet{ElBadry2024NS}. $^{c}$ The same 558 orbits: no MS root because $\mathcal A$ exceeds $\mathcal A_{\rm max}$ of a coeval MS companion, the luminous root excluded and the faint one allowed, or both roots excluded.}
\end{table}

We extend the test to 81\,981 \Gaia\ DR3 astrometric orbits selected from the triage catalogue of \citet{Shahaf2023}. We fit XP and 2MASS for each target, add \SPHEREx\ where available, and use the orbital parallax and a shared extinction for both stars. We retain converged fits with an acceptable $\chi^2$ and no parameter at the edge of its allowed range, obtaining companion-mass limits for 64\,691 systems.

We exclude a single MS companion in 2841 systems. Their minimum orbital companion masses exceed the largest MS companion masses allowed by the spectra by more than twice the catalogue mass error. Restricting the sample to the calibrated brightness and primary-mass range gives 1618 exclusions. Bootstrap draws of the calibration factors give a range of 2196--3652 for the full sample and 1302--2014 for the restricted one. The full-sample range does not include the uncertainty of extrapolating the threshold to the 1223 systems with primaries above 0.9\,\Msun, where we have no calibrating binaries.

When we fit the spectra and photocentre orbit jointly, the dark-companion model has a lower $-2\ln\mathcal L$ than the best luminous-companion model by more than 25 for 2754 of the 2841 systems.

The constraints weaken with distance. For primaries without detected companion light, the median companion-mass upper limit rises from 0.36\,\Msun\ within 100\,pc to 0.88\,\Msun\ at 500--1000\,pc. At these median limits, we can exclude a 0.6\,\Msun\ MS companion in the nearby sample, while the distant sample still allows it. We discuss the calibration range and its extension to more distant systems in Sect.~\ref{sec:limitations}.

\subsubsection{Comparison with the colour excess}\label{sec:colour-test}

An unresolved pair of low-mass stars can be redder than a single star of the same luminosity. Among systems whose orbits exclude a single MS companion, \citet{Shahaf2024,Shahaf2025} select compact-companion candidates that lack this colour excess. We correct the colour excess for the APOGEE or LAMOST \citep{Zhang2025DDPayne} metallicity using $-0.11$\,mag per dex and compare their selection with our spectral mass limits.

We test both methods on systems whose companions are independently known to be faint. \citet{Yamaguchi2024} combine RV and astrometric orbits to constrain companion masses and light, assigning their white-dwarf candidates to Group~1 and possible triples to Group~2. Among the 24 Group~1 systems with usable XPS limits, we select 22 with XPS, compared with 14 from the colour test. In a separate follow-up sample of \citet{ElBadry2026}, we select all 22 systems with usable limits whose RVs support a dark companion, while the colour test again selects 14.

In the corrected catalogue of \citet{Shahaf2025}, 553 systems have a colour excess ($\Pr(\mathrm{red})\ge64$ per cent) although XPS excludes a single MS companion. Of these, 35 per cent belong to class~III, where even an MS pair cannot explain the orbit.

We use orbital eccentricities to estimate the luminous stellar systems that may enter each selection. We take the systems that both methods classify as faint and those that both classify as luminous as pure templates for the two eccentricity distributions, then fit their mixture to the systems where the methods disagree. We assume that the templates represent the corresponding populations in these disagreement samples. We repeat the full-period fit over three ranges of $\log_{10}(P/{\rm d})$ (1.5--4, 2.3--2.7 and 2.6--2.95). Across these windows, the mixture gives an MS-like share of 4--5 per cent for the XPS selection and 12--18 per cent for the colour selection. The spectral mass limit recovers more of the independently supported dark companions than the colour selection. Under the assumptions of the mixture model, it also leaves a smaller MS-like component in the compact-companion sample.

\section{Discussion}\label{sec:discussion}

\subsection{The infrared signal of a companion}\label{sec:ir}

A cool companion contributes a larger fraction of the light in the infrared, so infrared spectra can help us identify companions. In mocks with measurement noise alone, we increase the median $\dchi$ fivefold by adding \SPHEREx\ at S/N$=20$ for $q\ge0.3$. For twins, the companion does not change the spectral shape, and we rely on the absolute luminosity to identify it.

For observed single stars, we find a scatter of about 2 per cent in the mean infrared flux around the model, comparable to the excess of some cool companions. When we inject model companions into the spectra of 458 observed single stars, adding $JHK_s$ and \SPHEREx\ to XP changes the AUC by less than 0.02 at every mass ratio (Table~\ref{tab:inject}). These differences between the observations and the model limit our ability to use weak infrared excesses. We can still see an infrared excess in individual binaries (Fig.~\ref{fig:outcomes}a), but adding \SPHEREx\ does not substantially improve the separation of binaries from single stars across the sample. For the 1714 literature binaries and 458 single stars at $9\le G<17.7$ without a resolved wide companion or a multi-peaked image, the AUC is 0.87 with XP alone and 0.86 with XP, $JHK_s$ and \SPHEREx. With the stricter bright limits of the injection test, it is 0.88.

As \SPHEREx\ accumulates more survey observations, repeated measurements will improve sensitivity and help test photometric consistency \citep{Bock2026}. Refining the internal calibration with survey data and the absolute flux calibration with external standard stars may further reduce systematic errors \citep{Akeson2025}. With more accurate single-star models, we may then make fuller use of the infrared signal of cool companions to improve binary detection.

\subsection{Limitations}\label{sec:limitations}

\textit{Models and sensitivity.} We validate the emulator on observed stars above 0.09\,\Msun. Below this mass, the theoretical models carry errors of 20--30 per cent in the infrared and exist at solar metallicity only, so the metallicity dependence of the emulator rests on the data-driven network alone. These errors can mimic or hide a companion, notably at the stellar--substellar transition and in the blue XP flux of cold dwarfs.

We adopt the metallicity scale of the PARSEC isochrones. For the orbits of Sect.~\ref{sec:nss-catalogue} with an APOGEE or LAMOST metallicity, we obtain median offsets from the spectroscopic values of +0.16\,dex for APOGEE and +0.20\,dex for LAMOST. For the triple \Gaia\ DR3 1748901959855337472, both fits reach the upper age bound of 10\,Gyr (Fig.~\ref{fig:triple}).

We obtain weaker companion constraints for lower-mass primaries. For primaries of 0.5--0.7\,\Msun\ with no detected companion light, the median $q_{\rm up}$ is 0.87, and below 0.5\,\Msun\ it is 1 (Sect.~\ref{sec:binary}). We therefore cannot exclude an equal-mass companion for a typical undetected primary below 0.5\,\Msun. Near the hydrogen-burning boundary, we also cannot exclude a low-mass star or an unresolved pair in individual systems.

\textit{Threshold calibration and applicability.} We use $D>25$ to select systems that favour the binary model, but we have not calibrated this threshold to a common statistical significance. For the mass-ratio profiles, we derive thresholds $t$ from mocks with observed single-star residuals and calibrate the scaling factors $K$ on observed binaries. Within each mass bin, we use one threshold over $9\le G<17.7$, without further separating stars by brightness. We do not calibrate the flux-ratio limit $\beta_{\rm up}$ separately. The jump in $K$ at $q=0.5$ also causes $q_{\rm up}$ to avoid 0.45--0.5 and accumulate just above 0.5.

Applying this calibration to all DR3 orbits extends it over a wider range of mass, distance and extinction. For systems outside the calibrated range, we adopt the threshold of the nearest mass bin. These include 22 per cent of the orbits in Fig.~\ref{fig:dark-companions}, 47 per cent of all DR3 orbits in Sect.~\ref{sec:nss-catalogue}, and all MS-excluded systems with a primary above 0.9\,\Msun. We calibrate on GCNS stars within 100\,pc without extinction, and the mock calibration sample contains no primary below $\mh=-0.4$. In contrast, most systems in the all-orbit sample lie at 200--1000\,pc, where extinction is no longer negligible. We also use orbital parallaxes without the zero-point correction \citep{ElBadry2026}, which makes the 10\,pc fluxes about 3 per cent too bright at 440\,pc.

\textit{Benchmark sample.} Unrecognised binaries in the controls lower the measured AUC and raise the apparent FPR of $D>25$. We also obtain mass ratios concentrated towards 0.6--0.8. Among stars below 0.5\,\Msun\ on the MS with $D>25$, we find independent binary signatures in only a fifth, no more than for a CMD cut selecting the same number of stars.

\textit{Orbital reliability and companion identity.} We combine spectral upper limits with orbital masses to exclude MS companions, so the results depend on the reliability of the orbits. RV follow-up by \citet{ElBadry2026} refutes three orbits with usable spectral limits, two of which would pass our mass cut using the published orbit. The spectral mass cut therefore does not establish the reliability of an astrometric orbit. In addition, four neutron-star candidates and Gaia BH2 and BH3 have no spectral upper limit that passes the fit cuts.

We exclude a single coeval MS companion. An unresolved inner pair of the same total mass can be fainter and can therefore still explain some systems. Our XPS fits also contain no white-dwarf component. We test white-dwarf models separately with \texttt{spbin} in Appendix~\ref{app:spbin-wd}. We can further test individual candidates through white-dwarf signatures in XP spectra \citep{Li2025WDMS} or an ultraviolet (UV) excess.

The eccentricity distribution provides a population test, while individual white-dwarf binaries can still have eccentric orbits. In the mixture model of Sect.~\ref{sec:colour-test}, we assume pure templates and the same component distributions in the agreement and disagreement samples. The MS-like shares we infer depend on these assumptions.

With \Gaia\ DR4 epoch astrometry, we can include companion light directly in the orbit fit and compare a dark companion, one MS star and an inner pair. For more distant systems, we also need to correct the parallax zero point and extend the calibration over the corresponding range of brightness and extinction.

\section{Conclusions}\label{sec:conclusions}

We compared single-star and coeval-binary fits to the absolute XPS spectra of about 71\,000 nearby low-mass stars. We use companion light to identify binaries and distinguish mass-ratio solutions of photocentre orbits, and constrain companion masses when we detect no companion light.

\begin{enumerate}
\item We identify binaries that CMD selection misses. These include systems whose companions add little light and those in which companion light lifts a metal-poor primary onto the MS ridge. At the same FPR, we recover 44 per cent of the known binaries the CMD cut misses and retain 86 per cent of those it selects. In the GCNS, independent binary signatures support the gain over CMD selection for primaries of 0.5--0.9\,\Msun. Below 0.5\,\Msun, we find no corresponding gain. In the benchmark sample, adding \SPHEREx\ does not improve the separation of binaries from field controls over XP alone.

\item We use companion light to distinguish the faint and luminous solutions of photocentre orbits. Of 487 systems with detected companion light, we distinguish the two solutions by more than 25 in $-2\ln\mathcal L$ for 257. The faint solution can assign a low mass to a companion nearly as massive as its primary, allowing stellar binaries to enter substellar candidate lists. Among candidates with RV follow-up, all 11 binaries identified from double or asymmetric lines have larger $D$ than any of the 16 confirmed brown-dwarf or planet hosts.

\item We combine spectral mass limits with orbital masses to identify companions fainter than a single MS star. When we detect no companion light, we set an upper mass limit for a coeval MS companion. If the minimum orbital mass exceeds this limit by more than twice its catalogue error, we exclude a single MS star of that mass. Of 24 systems with spectral limits and companions independently established as faint through RV studies, we select 22, compared with 14 selected by the absence of a colour excess. Among 64\,691 DR3 orbits with a clean fit and a limit, we select 2841 systems, including 1618 within the calibrated brightness and primary-mass range. Within 200\,pc, their companion masses cluster at 0.53--0.62\,\Msun, supporting a white-dwarf interpretation. We exclude lighter MS companions in fewer systems, so the lower edge of this range is partly set by our sensitivity. Unresolved inner pairs can also explain some systems.
\end{enumerate}

We provide the spectral and photocentre-orbit tests in the public package \texttt{spbin} (\url{https://github.com/jiadonglee/spbin}) for application to forthcoming \Gaia\ DR4 orbital solutions.

\begin{acknowledgements}
This work uses data from the European Space Agency mission \Gaia, processed by the \Gaia\ Data Processing and Analysis Consortium. It uses \SPHEREx\ observations and services of the NASA/IPAC Infrared Science Archive, as well as data products from 2MASS and the unWISE catalogue. \SPHEREx\ is funded by NASA and is a joint project of the Jet Propulsion Laboratory and the California Institute of Technology. 
We thank C. Hainje and D. Hogg for their work on the SPHEREx data-server pipeline.
We thank F. Davies and E. Ba\~nados for the \SPHEREx\ extraction code \texttt{spherextract} and  \texttt{spherex-tools}, on which its download of the image cutouts is based. 
During the preparation of this work, the authors used the AI assistants Claude Opus 5.5 (Anthropic) and GPT-6-Astra (OpenAI) to assist with language editing, and revision, as well as code development, review, and testing. 
All AI-assisted outputs were reviewed and validated by the authors.
JL acknowledges the European Research Council for support from the ERC Advanced Grant ERC-2021-ADG101054731.
\end{acknowledgements}

\section*{Data availability}
The catalogues of the XPS fits and companion limits, the J-CaPS XPS emulator and its calibration files are available at Zenodo (\url{https://doi.org/10.5281/zenodo.24319325}). The package \texttt{spbin} is available at \url{https://github.com/jiadonglee/spbin}.

\bibliographystyle{aa}
\bibliography{references}

\begin{appendix}

\section{Training of the data-driven network}\label{app:training}

We use a multilayer perceptron with two hidden layers of 16 units and Gaussian error linear unit (GELU) activations. The network has 3192 trainable parameters and maps $(\teff,\GKs,\feh)$ to the spectral shape $r_\lambda$ in 168 channels: 61 XP samples, five broad bands and 102 \SPHEREx\ channels. We convert this output to absolute fluxes using \MKs\ through Eq.~\eqref{eq:anchor}. Table~\ref{tab:labels} lists the sources and ranges of the training labels. We derive \MKs\ and \GKs\ from each star's photometry and parallax. The evolutionary temperatures of the ultracool dwarfs do not depend on \SPHEREx\ spectra.

We use a Cauchy loss on the linear-flux residuals and add a penalty on the curvature of the predicted spectra. During training, we assign weights of 0.4, 0.4 and 0.2 to the cool, warm and ultracool groups. For each ultracool dwarf with unknown metallicity, we marginalise over five metallicity nodes from $-0.5$ to $+0.5$ with a Gaussian prior of width 0.3\,dex. We train with a learning rate of $3\times10^{-4}$ and a weight decay of $10^{-5}$, and select the final network using the loss on a held-out development set. We give the three groups equal weight in this evaluation.

\begin{figure*}
\centering
\includegraphics[width=\textwidth]{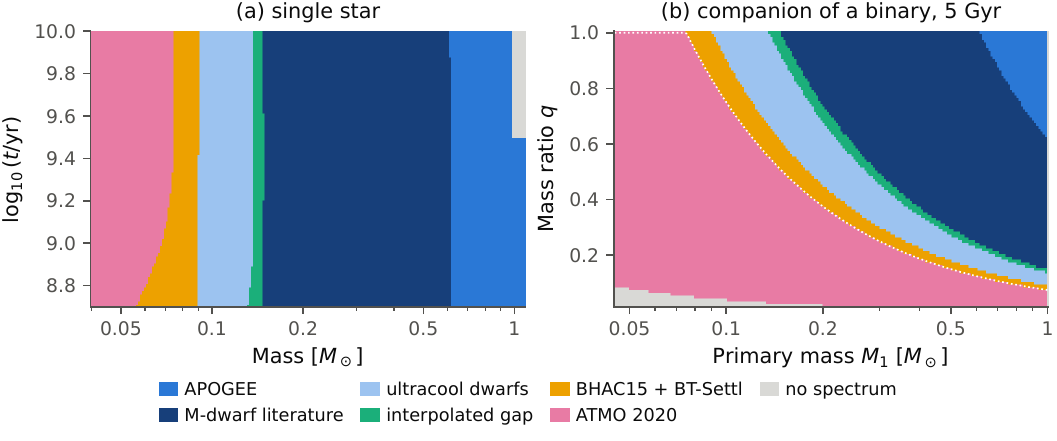}
\caption{Source of the emulated spectra at solar metallicity, (a) for a single star and (b) for the companion of a primary $M_1$ at 5\,Gyr (dotted: 0.075\,\Msun). Blue shades: data-driven network, by training labels (APOGEE, literature M dwarfs, ultracool dwarfs); green: interpolation (0.136--0.147\,\Msun); yellow: BHAC15 + BT-Settl; pink: ATMO\,2020; grey: no spectra.}
\label{fig:routes}
\end{figure*}

Figure~\ref{fig:routes} shows which model provides the spectrum at each mass, for a single star and for the companion of a given primary.

After training, we apply two corrections to the model output. First, we fit a smooth near-infrared correction with 18 coefficients to the training stars to adjust the \SPHEREx\ model spectra of cool stars, with a temperature taper from 2800 to 3400\,K. Second, we adjust the XP and \SPHEREx\ output layers at 0.88--1.12$\mum$, with a three-per-cent soft constraint on their agreement at 0.94--0.992$\mum$. The latter reduces the median disagreement in this interval to 0.06 per cent for cool stars and 0.44 per cent for warm stars. Across the wider overlap at 0.75--0.99$\mum$, the median disagreement remains 1--6 per cent, with the largest differences at the blue edge of the \SPHEREx\ range. We fit both corrections using only training stars and retain the original observed fluxes and their uncertainties.

We also test interpolation by removing intervals from the data-driven model. Interpolation across these gaps reproduces the removed spectra to better than 0.2 per cent.

We estimate the model covariance in Eq.~\eqref{eq:cmodel} from validation-star data with S/N$\,>5$. We compute the second moment of the log residuals $\ln(F_{\rm obs}/F_{\rm mod})$ about zero, so that any mean model offset contributes to the error, and subtract the mean measurement variance of each channel from the diagonal. We retain the four largest positive eigenvalues and their eigenvectors, and describe the remaining variance with a non-negative diagonal term. For ATMO\,2020, we use the observed residuals of the 51 brown dwarfs in Appendix~\ref{app:physical}. For BHAC15 + BT-Settl, we use its differences from the data-driven model where both are defined. We retain three eigen-directions for each set of theoretical models. Because we fix the $K_s$ flux through the input label \MKs, we set the data-driven model error in this channel to zero.

\begin{figure*}
\centering
\includegraphics[width=\textwidth]{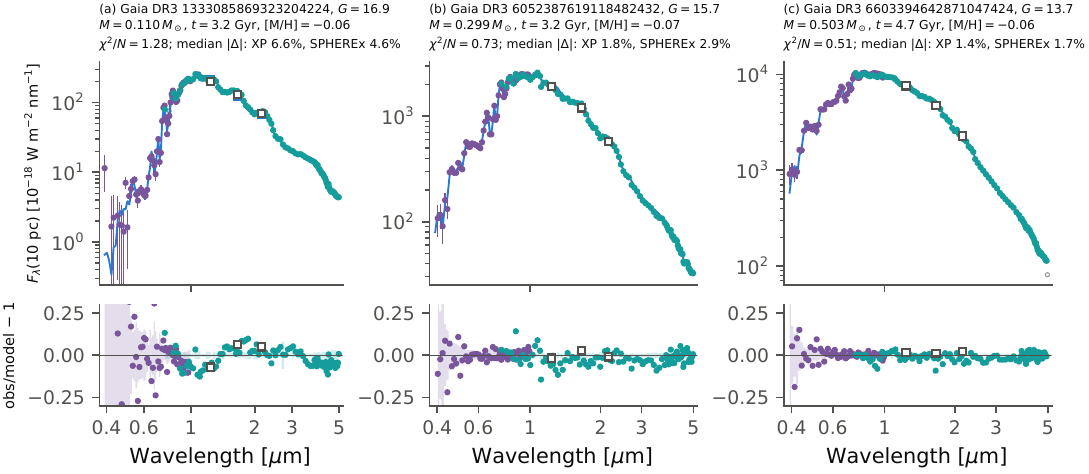}
\caption{Single-star fits of three GCNS stars at 0.1--0.5\,\Msun: XP (purple), \SPHEREx\ (teal), $JHK_s$ (grey squares), masked channels (open grey) and the single-star model (blue), with fractional residuals below (shaded: measurement uncertainty).}
\label{fig:fits}
\end{figure*}

Figure~\ref{fig:fits} shows fits to GCNS stars near 0.1, 0.3 and 0.5\,\Msun. We fit XP, $JHK_s$ and \SPHEREx\ together using the continuous single-star fits of Appendix~\ref{sec:gcns-likelihood}. We select stars near the MS ridge with $D<2$, $G<17$ and at least 90 fitted \SPHEREx\ channels. For each target mass, we retain stars with fitted masses within 0.01\,\Msun\ and show the one whose $\chi^2$ per fitted channel is closest to the median of that group.

\FloatBarrier
\section{Test of the theoretical models}\label{app:physical}

We test the BT-Settl and ATMO\,2020 atmospheres against \SPHEREx\ spectra of field brown dwarfs. We select known brown dwarfs from the public binned spectra of the SPIFF library \citep[][see also the brown-dwarf sequence of \citealt{Rustamkulov2026}]{Gagne2026}, requiring \Gaia\ and $K_s$ photometry, literature evolutionary temperatures of 1600--2000\,K and the catalogue's field flag. We retain the existing sample split and compare the models with its 13 validation objects. We adopt the temperatures, surface gravities and luminosities of \citet{Sanghi2023} and compute the radii as $R/R_\odot=\sqrt{L/L_\odot}\,(5772\,\mathrm K/\teff)^2$.

For each object, we compute BT-Settl and ATMO\,2020 spectra at the same fixed temperature, surface gravity and radius to compare the two atmospheres at the same stellar parameters. We integrate the model spectra over Gaussian channels on the observed wavelength grid and convert the observed spectra to absolute fluxes at 10\,pc using Eq.~\eqref{eq:f10}. We compare the absolute fluxes directly, without normalisation or a fitted amplitude.

Table~\ref{tab:physical} gives the median over the 13 objects of their median absolute fractional residuals. BT-Settl gives residuals of 17.4 and 20.5 per cent at 1--2 and 2--5$\mum$, respectively, compared with 23.1 and 28.4 per cent for ATMO\,2020. At 2--5$\mum$, BT-Settl has a median signed offset near zero, although it does not reproduce the flux in each channel accurately. ATMO\,2020 is systematically faint by a median of 27 per cent. At these fixed literature parameters, the two models show infrared residuals of about 20--30 per cent for brown dwarfs at 1600--2000\,K.

We also use these 13 validation objects, together with the 38 training objects, to estimate the ATMO\,2020 model covariance in Appendix~\ref{app:training}. This comparison therefore measures the model--data differences but does not independently test that covariance.

\begin{table}
\caption{Median per-object residual of the theoretical-model atmospheres on 13 field brown dwarfs at 1600--2000\,K.}\label{tab:physical}
\centering
\small
\begin{tabular}{lrr}
\toprule
Atmosphere & 1--2$\mum$ & 2--5$\mum$ \\
\midrule
BT-Settl & 17.4\% & 20.5\% \\
ATMO\,2020 & 23.1\% & 28.4\% \\
\bottomrule
\end{tabular}
\tablefoot{Median over objects of the per-object median absolute fractional residual in channels with S/N$>5$, at fixed literature temperature, gravity and radius.}
\end{table}

\FloatBarrier
\section{Mass-ratio sensitivity in mock spectra and companion injections}\label{app:mocks}

We first use mock spectra with measurement noise to measure how the companion signal varies with mass ratio and S/N. We then add model companions to observed single-star spectra to test how much the infrared data improve detection in the presence of the residuals in real spectra.

\subsection{Mock observations}\label{sec:grid}

We vary eight primary masses (0.06--0.95\,\Msun), five mass ratios ($q=0.1$, 0.3, 0.5, 0.7 and 1), three ages (1, 5 and 9\,Gyr) and three metallicities. We also include single stars. Of the 432 configurations, 208 binaries and 55 single stars lie within the emulator range. We use non-solar metallicities only within the range of the data-driven model.

To study the dependence on S/N, we fix the age at 5\,Gyr and the metallicity at solar. For the same eight primary masses, we vary $q$ from 0.1 to 1 in steps of 0.025. This gives 296 binaries, alongside the single stars, at each S/N from 5 to 100.

For each configuration we draw one set of standard Gaussian deviates and scale them by $\sigma_\lambda=F_\lambda/(\mathrm{S/N})$ at each S/N. We add this noise to the model fluxes and fit the same mock observations with both models. We use the measurement errors to compute $\chi^2$ and denote the best single-star value minus the best binary value by \dchi.

\subsection{Grid fits and the benchmark}\label{sec:gridfits}

We search 173 masses from 0.0053 to 1.05\,\Msun, 14 ages from 0.5 to 9.9\,Gyr and three metallicities ($-0.5$, 0 and $+0.3$). For binaries we include $q=0.02$ and scan $q=0.05$ to 1 in steps of 0.05, or 0.025 for the S/N scan. We evaluate every template, so the fitted $q$ takes one of the grid values.

For the benchmark stars we follow Sect.~\ref{sec:chi2}, using a finer grid with mass steps of 1 per cent up to 1.1\,\Msun\ and metallicity steps of 0.1\,dex from $-1.0$ to $+0.5$. At non-solar metallicity we use solar-metallicity theoretical spectra for companions below the range of the data-driven model. At each $q$ we select the eight best templates using the diagonal of $\mathbf C$ and refit them with the full covariance. As in the mock tests, we use the same observations for both hypotheses and omit $JHK_s$, $W1$ and $W2$.

We mask isolated \SPHEREx\ channels with fluxes above ten times the median of their neighbours. We require $G\ge8$ because brighter stars have saturated or outlying channels. We match the controls in apparent magnitude and colour. Matching in absolute magnitude would preferentially select binaries as controls for over-luminous systems.

Two DEBCat controls at $G\approx8.2$ have single-star $\chi^2$ above 100 per \SPHEREx\ channel. We retain both. The DEBCat AUC with \SPHEREx\ is 0.82 with these controls and 0.92 without them.

\begin{figure*}
\centering
\includegraphics[width=\textwidth]{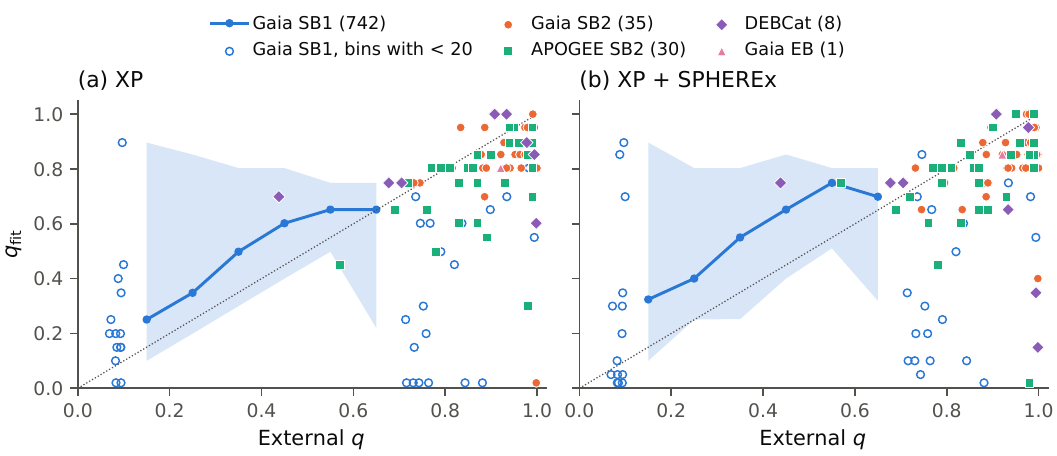}
\caption{Fitted against external mass ratio for the 816 benchmark binaries, grid fits with XP alone (a) and with \SPHEREx\ (b). \Gaia\ SB1: medians and 16--84 per cent ranges in bins of 0.1 that hold at least 20 binaries, open circles elsewhere. SB2 and eclipsing systems are shown individually. The dotted line is one-to-one.}
\label{fig:bench}
\end{figure*}

Figure~\ref{fig:bench} compares the fitted mass ratios of the benchmark binaries with the external values.

\subsection{Companion signals in mock spectra}\label{sec:landscape}

\begin{figure*}
\centering
\includegraphics[width=\textwidth]{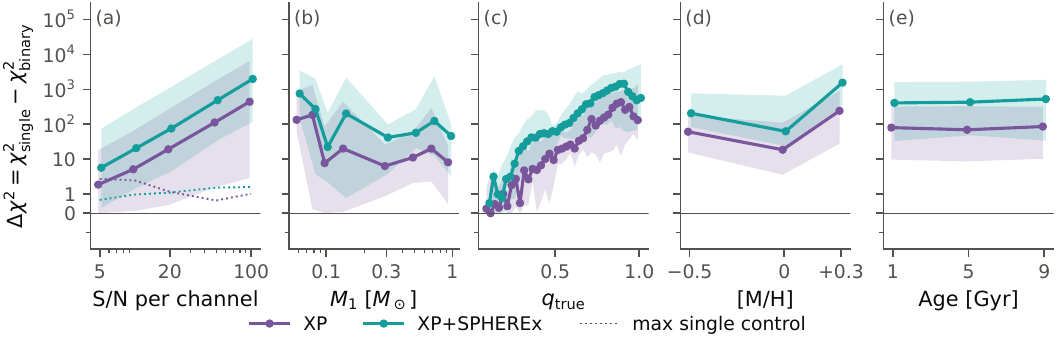}
\caption{\dchi\ of mock binaries with XP alone (purple) and with \SPHEREx\ (teal): medians and 16--84 per cent ranges against S/N (a; dotted: largest single-star value), and at S/N$=20$ against $M_1$, $q$, metallicity and age (b--e). The $y$ axis is linear below 1.}
\label{fig:dchi2}
\end{figure*}

Adding \SPHEREx\ raises the median \dchi\ from 1.8 to 5.6 at S/N$=5$, from 19 to 75 at S/N$=20$ and from 442 to 1987 at S/N$=100$ (Fig.~\ref{fig:dchi2}a). In both fits \dchi\ grows approximately as $(\mathrm{S/N})^2$. No mock single star in the S/N scan exceeds $\dchi=3$.

At fixed S/N the signal depends most strongly on mass ratio. At S/N$=20$, adding \SPHEREx\ raises the median \dchi\ from 0.5 to 2.4 for $q<0.3$ and from 29 to 143 for $q\ge0.3$. The signal rises by more than three orders of magnitude from $q=0.1$ to $q\approx0.9$, then falls towards $q=1$ (Sect.~\ref{sec:discussion}). Across primary masses the median varies by an order of magnitude with no clear trend.

We compare ages and metallicities for the ten $(M_1,q)$ pairs that the grid covers at every age and metallicity (Fig.~\ref{fig:dchi2}d,e). For these pairs \dchi\ depends little on age. It is lowest at solar metallicity and up to 25 times higher at $\mh=+0.3$.

\subsection{Companions injected into observed single stars}\label{sec:inject}

To test how much the infrared data help in observed spectra, we use 458 literature single stars with $9\le G<17.7$ and no resolved wide companion or multi-peaked image (Appendix~\ref{sec:lit-details}). For each star we use the parameters of its single-star XP fit to compute coeval companion spectra at $q=0.2$, 0.3, 0.4 and 0.6. We add each companion spectrum, including photon noise on its flux, to the original observations. This preserves the observed residuals, missing channels and measurement errors. We run the continuous fits of Appendix~\ref{sec:gcns-likelihood} with XP alone and with XP, $JHK_s$ and \SPHEREx.

Table~\ref{tab:inject} gives the AUC at each injected mass ratio. The largest gain from adding the infrared data occurs at $q=0.6$ and comes mainly from primaries below 0.5\,\Msun.

\begin{table}
\caption{Companions injected into 458 observed single stars.}\label{tab:inject}
\centering
\small
\begin{tabular}{lrr}
\toprule
$q$ & XP & XP, $JHK_s$, \SPHEREx \\
\midrule
0.2 & 0.720 & 0.701 \\
0.3 & 0.862 & 0.856 \\
0.4 & 0.912 & 0.914 \\
0.6 & 0.927 & 0.945 \\
\bottomrule
\end{tabular}
\tablefoot{AUC of $D$. These fits leave out \SPHEREx\ channels brighter than AB magnitude 10.5 at 2.42--3.82$\mum$ and 9.25 at other wavelengths.}
\end{table}

\FloatBarrier
\section{Mass-ratio profiles and their calibration}\label{app:mock}

\subsection{Mass ratios allowed by the spectra}\label{sec:gcns-likelihood}

Even when we do not detect a companion, we can ask which mass ratios are consistent with the observed spectrum. For each GCNS star we fit a single star and a binary with free $q$ to XP, $JHK_s$ and \SPHEREx. We start the continuous fits from the best grid template and omit isolated outliers and \SPHEREx\ channels brighter than AB magnitude 9. We allow the flux scale to vary with a Gaussian prior of width $\sigma_s$, equivalent to first order to the flux-scale term in Eq.~\eqref{eq:lnl}.

\begin{figure*}
\centering
\includegraphics[width=\textwidth]{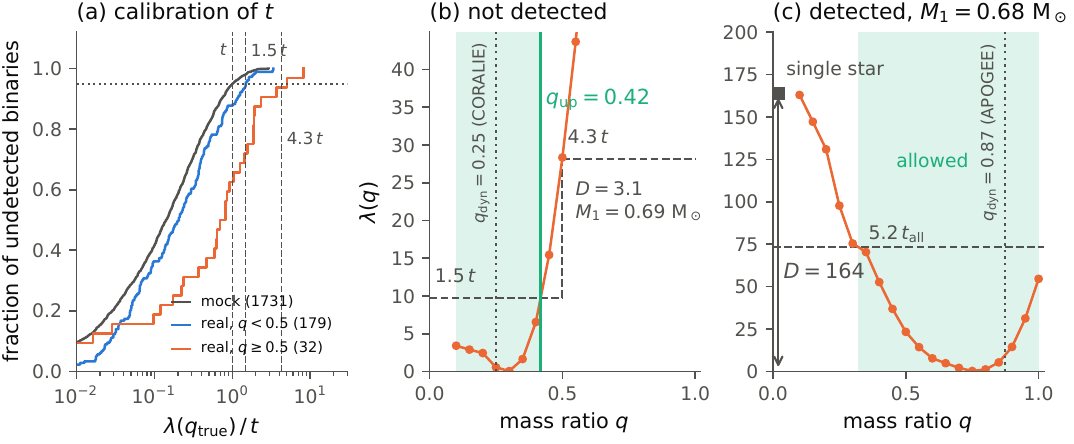}
\caption{Calibration and use of the mass-ratio profiles. (a) Cumulative distribution of $\lambda$ at the true mass ratio in units of $t$, for the mock binaries with $D<25$ (black), whose 95th percentile (dotted) defines $t$ in each primary-mass bin, and for observed undetected binaries with a dynamical mass ratio and a companion below 0.45\,\Msun, at $q<0.5$ (blue) and $q\ge0.5$ (orange), whose 95th percentiles give $K_{\rm f}$ and $K_{\rm l}$. (b) Profile of an undetected binary with a dynamical mass ratio (dotted): the threshold $K(q)\,t$ (dashed) gives $q_{\rm up}$. (c) Profile of a detected binary: the single star lies $D$ above the best model, and the allowed mass ratios (shaded) are those below $5.2\,t_{\rm all}$.}
\label{fig:lambda}
\end{figure*}

We then fix $q$ at each of 19 values from 0.10 to 1.00 and refit the primary mass, age, metallicity and flux scale. We call the smallest $-2\ln\mathcal L$ among the single-star fit, the free binary fit and the 19 fixed-$q$ fits $S_{\rm best}$. Fixing the mass ratio at $q$ increases the fit statistic by
\begin{equation}\label{eq:lambda}
 \lambda(q)=-2\ln\mathcal L(q)-S_{\rm best}.
\end{equation}
For a threshold $t$ we retain all mass ratios with $\lambda(q)\le t$ and define
\[
 q_{\rm up}=\max\{q:\lambda(q)\le t\}.
\]
We interpolate linearly where the profile crosses the threshold between adjacent nodes. Figure~\ref{fig:lambda}b,c shows the profiles of two GCNS binaries with dynamical mass ratios. For an M dwarf the spectrum can allow an equal-mass companion while excluding intermediate mass ratios (Sect.~\ref{sec:binary}). The upper limit alone therefore does not specify the full set of allowed mass ratios.

\begin{figure*}
\centering
\includegraphics[width=\textwidth]{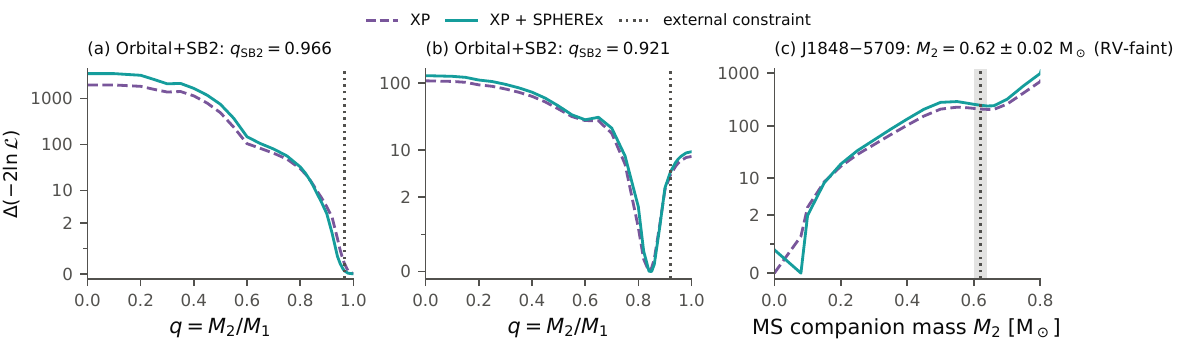}
\caption{Profiles of three systems with an external constraint (dotted), with XP alone (purple dashed) and with \SPHEREx\ (teal): $\Delta(-2\ln\mathcal L)$ against $q$ for two \Gaia\ \texttt{Orbital+SB2} systems (a, b) and against the MS companion mass for J1848$-$5709 \citep[c;][shaded: uncertainty of the companion mass]{Yamaguchi2024}, whose spectra are shown in Fig.~\ref{fig:outcomes}c.}
\label{fig:nss-truth}
\end{figure*}

We test the profiles against independent orbital constraints for three systems (Fig.~\ref{fig:nss-truth}). For two \Gaia\ \texttt{Orbital+SB2} binaries, double-lined spectroscopy gives $q=0.966$ and 0.921, and our profiles reach their minima at 1.00 and 0.85, respectively. Fixing $q$ at the spectroscopic values raises $-2\ln\mathcal L$ by only 0.1 and 4.7, whereas the single-star fits lie 3477 and 129 above the minima.

For J1848$-$5709, fixing an MS companion at the orbital mass of $0.62\pm0.02$\,\Msun\ \citep{Yamaguchi2024} raises $-2\ln\mathcal L$ by 246 above the minimum. Adding \SPHEREx\ has little effect on the low-mass side of the minimum but steepens the profile at higher companion masses.

\subsection{Calibration with mock and observed binaries}\label{sec:calibration}

We seek a threshold that includes the true mass ratio among the allowed values for about 95 per cent of binaries. With the usual single-parameter Gaussian threshold of 3.84, the allowed set contains the true $q$ for only 44--93 per cent of undetected mock binaries with observed residuals, depending on primary mass. Model errors make the profiles too narrow, and the usual $\chi^2$ approximation breaks down near $q=0$, where the binary approaches a single star. We first estimate a threshold for each primary-mass bin using mock binaries, then calibrate it with observed binaries that have independent mass ratios.

We group reference stars by single-star fitted mass, with bin edges at 0.1, 0.2, 0.3, 0.5, 0.7 and 0.9\,\Msun. We select stars within $-0.2$ to $+0.5$\,mag of the MS ridge with a white-dwarf probability below 0.05 and no \Gaia\ non-single-star solution, resolved pair or multi-peak flag. Within each mass bin we divide the stars by $G$ at 9, 11, 13, 15 and 17.7\,mag. We draw 800 examples from each mass and magnitude bin containing at least 300 reference stars, giving 8800 mock systems.

For each system we start from its single-star fit and add a coeval companion with probability 0.35, drawing $q$ uniformly from 0.1 to 1. To include residuals typical of observed spectra, we choose another reference star in the same bin with the same model and similar $G$. In each channel we multiply the mock flux by that star's observed-to-model flux ratio and adopt its fractional uncertainty. We retain the original system's parallax and fit each mock spectrum in the same way as the observations.

For undetected systems we take the 95th percentile of $\lambda(q_{\rm true})$ among mock binaries with $D<25$ in each mass bin. This threshold $t$ is 6.6 in both bins spanning 0.5--0.9\,\Msun\ and reaches 27.9 at 0.1--0.2\,\Msun\ (Table~\ref{tab:thresholds}). The coverage averages over the uniform distribution of simulated $q$. Whenever the allowed set contains $q_{\rm true}$, its upper limit satisfies $q_{\rm true}\le q_{\rm up}$, so the upper limit covers the true mass ratio at least as often as the full set does.

We next test $t$ on observed binaries. We use the same companion model to generate and fit the injections, whereas a real companion can differ from its model spectrum. Its contribution to the residuals increases with its share of the light. The threshold $t$ excludes the measured $q$ for 11 of the 33 binaries with dynamical mass ratios and $D<25$ in Appendix~\ref{sec:lit-details}. It also excludes the measured $q$ for 13 per cent of 178 \Gaia\ \texttt{AstroSpectroSB1} systems with $D<25$ and companion masses below 0.45\,\Msun. We apply this mass cut to favour MS companions because few white dwarfs have lower masses.

We combine the two observed samples to calibrate separate factors for low and high mass ratios. The 179 systems with $q<0.5$, almost all \texttt{AstroSpectroSB1}, give $K_{\rm f}=1.5$ (16--84 per cent bootstrap range 1.3--1.6). The 32 systems with $q\ge0.5$, almost all from the dynamical mass-ratio sample, give $K_{\rm l}=4.3$ (2.7--6.4; Fig.~\ref{fig:lambda}a). For undetected companions we retain mass ratios satisfying $\lambda(q)\le K(q)\,t$. For detected systems we derive $t_{\rm all}$ from all mock binaries and calibrate it with detected binaries with dynamical mass ratios, giving a threshold of $5.2\,t_{\rm all}$. When we calibrate on one half of each observed sample and test on the other, the allowed sets contain the measured $q$ for 94--95 per cent of the test binaries.

The mocks also show how companion light affects the stellar parameters of a single-star fit. The median ratio of fitted mass to true primary mass rises from 1.02 at $q<0.5$ to 1.13 at $0.5\le q<0.8$ and 1.23 at $q\ge0.8$. The single-star fit also overestimates the metallicity by 0.2\,dex.

\begin{table}
\caption{Profile thresholds of the calibration bins.}\label{tab:thresholds}
\centering
\small\setlength{\tabcolsep}{4pt}
\begin{tabular}{lrrr}
\toprule
$M_1$ [\Msun] & $t$ & $t_{\rm all}$ & $5.2\,t_{\rm all}$ \\
\midrule
0.1--0.2 & 27.9 & 39.8 & 207 \\
0.2--0.3 & 20.7 & 24.8 & 129 \\
0.3--0.5 & 9.5 & 12.0 & 63 \\
0.5--0.7 & 6.6 & 14.1 & 73 \\
0.7--0.9 & 6.6 & 20.3 & 105 \\
\bottomrule
\end{tabular}
\tablefoot{One bin per range of single-star primary mass over $9\le G<17.7$. The thresholds $t$ and $t_{\rm all}$ are defined in the text.}
\end{table}

\FloatBarrier
\section{\texorpdfstring{Details of the analyses of the GCNS and the \Gaia\ DR3 orbits}{Details of the analyses of the GCNS and the Gaia DR3 orbits}}\label{app:details}

\subsection{Sample and fits}

We use \texttt{spherextract} to measure the stellar flux in each \SPHEREx\ exposure. We extract a cutout at the stellar position and sample the point-spread function on the detector pixels following the SPIFF pipeline \citep{Gagne2026}. After masking flagged pixels, we estimate the stellar flux with an inverse-variance-weighted matched filter, fit a local linear background and iteratively reject outlying pixels \citep{Horne1986}. We convert surface brightness to flux density using the solid angle of each pixel and combine the exposures into 102 spectral channels. We retain the observed flux zero point.

We convert the 2MASS $J,H,K_s$ and unWISE $W1,W2$ magnitudes to $F_\lambda$ using the Vega zero points of \citet{Cohen2003} and \citet{Jarrett2011}, respectively. For each star we take the median S/N over the fitted \SPHEREx\ channels. Across the sample, the median of these values is 84 and the 16--84 per cent range is 48--154. The typical S/N per XP sample is 46.

We select 70\,933 stars from the GCNS parent sample of 143\,523 stars (Sect.~\ref{sec:gcns-sample}). Table~\ref{tab:selection} gives the number remaining after each cut. We require $9\le G<17.7$ because brighter stars have few usable \SPHEREx\ channels and fainter stars have noisy XP spectra. We remove sources with multi-peaked images to reduce photometric blending. We measure the MS ridge in bins of 0.1\,mag in $BP-RP$ and remove stars more than 0.5\,mag fainter than the ridge at their colour.

\begin{table}
\caption{Selection of the GCNS sample.}\label{tab:selection}
\centering
\small\setlength{\tabcolsep}{4pt}
\begin{tabular}{lr}
\toprule
Step & Remaining \\
\midrule
Parent sample (Sect.~\ref{sec:gcns-sample}) & 143\,523 \\
$9\le G<17.7$ & 124\,162 \\
Single-star primary mass 0.1--0.9\,\Msun & 119\,409 \\
Not in a GCNS wide binary & 100\,933 \\
\texttt{ipd\_frac\_multi\_peak}$\,=0$ & 80\,796 \\
White-dwarf probability $<0.05$ & 78\,638 \\
\MKs\ available & 78\,155 \\
At most 0.5\,mag below the MS ridge & 70\,933 \\
\bottomrule
\end{tabular}
\end{table}

To set the bright limit for \SPHEREx, we fit 1671 reference MS single stars at $G=5$--11 using XP and $JHK_s$ alone and compare the predicted and observed \SPHEREx\ fluxes. At AB magnitudes 8.5--9 the observed fluxes are still systematically high, with median $\ln(F_{\rm obs}/F_{\rm model})$ values of $+0.02$ to $+0.09$. We therefore omit channels brighter than AB magnitude 9 from the continuous fits.

We also compare the log flux in each channel with the median of the three valid channels on either side. We flag a channel if the absolute difference exceeds both 0.2 and $5\sigma$. If more than ten channels are outliers, we omit the entire \SPHEREx\ spectrum from the fit.

\subsection{Literature binaries and single stars}\label{sec:lit-details}

We select 2947 known binaries from the literature for Sect.~\ref{sec:gcns}. \Gaia\ DR3 provides 1996 spectroscopic or astrometric--spectroscopic orbits. A further 556 systems have dynamical mass ratios from DEBCat, SB9 \citep{Pourbaix2004}, \Gaia\ DR3, APOGEE \citep{Kounkel2021} and other published orbits. We select 134 SB9 SB1 systems and 155 APOGEE binaries \citep{PriceWhelan2020} whose mass functions require a companion of at least 0.08\,\Msun. The rest are 72 \Gaia\ DR3 eclipsing binaries \citep{Mowlavi2023} and 34 stellar companions from RV follow-up of \Gaia\ astrometric candidates (Sect.~\ref{sec:gcns-orbits}), the California Legacy Survey \citep{Rosenthal2021} and CARMENES \citep{Ribas2023}.

We use 2418 stars with no stellar companion detected by RV monitoring as controls. These include 2025 \Gaia\ RV standards with a scatter below 0.1\,km\,s$^{-1}$ over at least 1000\,d \citep{Soubiran2018} and 104 APOGEE stars with at least six observations spanning at least a year and velocities consistent with a constant. The remaining controls are 174 CARMENES M dwarfs, 101 California Legacy Survey stars and 14 hosts of confirmed substellar companions.

Within the GCNS magnitude range ($9\le G<17.7$), we have 2266 literature binaries and 708 controls. We use these controls to estimate the false-positive rate for the GCNS selection (Table~\ref{tab:literature}). For the 2172 stars in the infrared comparison (Sect.~\ref{sec:ir}), we also check for \Gaia\ neighbours within 6\arcsec\ with $G_{\rm neighbour}\le G_{\rm target}+2$ \citep{Rybizki2022}. Excluding the 17 targets with such a neighbour leaves the AUC of $D$ at 0.87.

\subsection{Photocentre orbits}\label{sec:photocentre-details}

We use the two-body parallax to scale the observed fluxes and uncertainties to 10\,pc. A single-star astrometric fit absorbs part of the orbital motion. The single-star and orbital parallaxes differ by more than 5 per cent for 25 per cent of the \texttt{Orbital} systems and 35 per cent of the \texttt{AstroSpectroSB1} systems. We compute $\beta_G(q)$ and find the orbital roots over $0.1\le q\le1$ using the primary mass, age and metallicity of the free binary fit. For $q<0.1$, outside the emulator range, we take $\beta_G=0$ and use the single-star likelihood for the faint root $q_{\rm dark}$. Recomputing the roots at the refitted primary parameters changes $q$ by a median of 0.007.

We test the roots with 11 \Gaia\ \texttt{Orbital+SB2} systems and four APOGEE SB2 systems \citep{Kounkel2021}. We derive the \Gaia\ spectroscopic mass ratios from $K_1/K_2$, taking the reciprocal for the six ratios above unity. For the one APOGEE system with an \texttt{AstroSpectroSB1} solution, we use only the photocentre orbit. In seven of the nine \Gaia\ systems with a clear choice of root, adding the orbit brings the fitted mass ratio closer to the SB2 value. The largest discrepancy falls from 0.09 to 0.045. Table~\ref{tab:individual} lists the three systems with an ambiguous root or a disagreement with the SB2 mass ratio.

For \texttt{AstroSpectroSB1} systems we obtain $a_1\sin i$ from the RV orbit and use the astrometric inclination $i$ to recover the primary semimajor axis $a_1$. We solve the mass function for $q=q_{\rm dyn}$ at the primary mass of the free binary fit. From the photocentre semimajor axis $a_0$, we obtain the relative semimajor axis $a$, the companion's fraction of the total $G$-band light $\ell$ and its flux ratio to the primary $\beta_{\rm dyn}$ as
\[
a=\frac{a_1(1+q)}{q},\qquad \ell=\frac{q}{1+q}-\frac{a_0}{a},\qquad \beta_{\rm dyn}=\frac{\ell}{1-\ell}.
\]
We propagate the uncertainties in the Thiele--Innes elements and the parallax and adopt a 10 per cent uncertainty in primary mass. The median uncertainty in $\beta_{\rm dyn}$ is 0.028, so these orbits typically cannot detect flux ratios below about 0.05. In the two bins $q_{\rm dyn}=0.3$--0.5 and 0.5--0.7, the median orbital flux ratios are 0.012 and 0.042, compared with model predictions of 0.016 and 0.044. In each bin the prediction lies within the 16--84 per cent interval of $\beta_{\rm dyn}$ for 55--57 per cent of systems.

\begin{figure*}
\centering
\includegraphics[width=\textwidth]{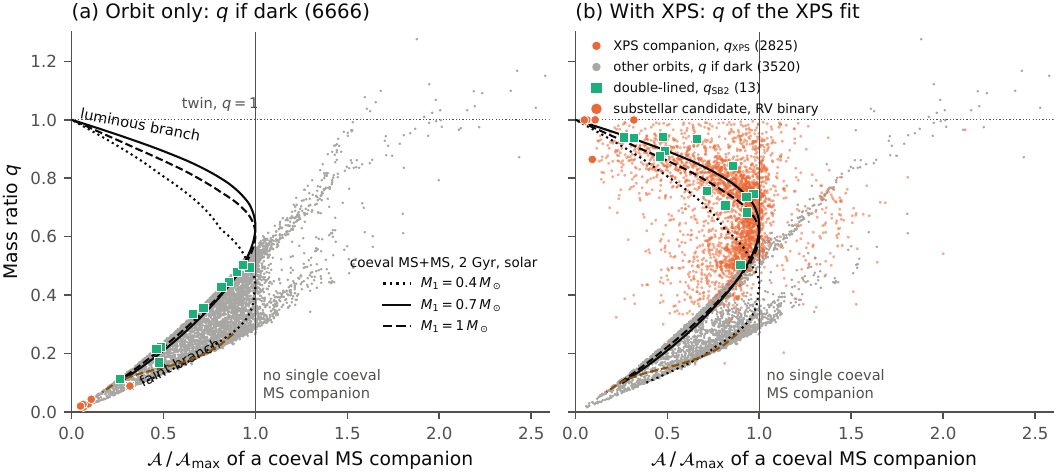}
\caption{\Gaia\ DR3 orbits within 200\,pc against $x=\mathcal A/\mathcal A_{\rm max}$, where $\mathcal A_{\rm max}$ is the largest $\mathcal A$ that a coeval MS companion gives for the primary of the single-star fit at its age and \mh. Black lines: $q(x)$ of coeval MS pairs for primaries of 0.4, 0.7 and 1\,\Msun\ (2\,Gyr, solar metallicity); brown dashed: a dark companion of 0.08\,\Msun. (a) Every orbit at its dark-companion mass ratio $q_{\rm dark}$. (b) Orbits with $D>25$ and an MS companion not excluded (orange) at $q_{\rm XPS}$ (fits for which the scan allows a twin companion omitted), the other orbits (grey) at $q_{\rm dark}$. Green squares: SB2 systems, at $q_{\rm dark}$ in (a) and at the SB2 mass ratio in (b); orange circles: substellar candidates whose RVs show a binary.}
\label{fig:orbit-branches}
\end{figure*}

For the orbits within 200\,pc (Fig.~\ref{fig:orbit-branches}), we group primaries in bins of 0.05\,\Msun. The 90th percentile of $\mathcal A$ in each bin corresponds to $\mathcal A/\mathcal A_{\rm max}=0.96$--1.10. The 50 \texttt{Orbital} solutions that favour the faint root have $q_{\rm XPS,dyn}=0.24$--0.49. Of these, \citet{Shahaf2023} classified 32 and assigned none a probability above 0.5 for either class~II or class~III.

\subsection{Upper limits on companion mass and flux ratio}\label{sec:limits-details}

We scan companion mass for 1782 GCNS orbits within 100\,pc and 4884 additional orbits with two-body parallaxes above 5\,mas. We apply the same test to the 26 Group~1 and five Group~2 systems of \citet{Yamaguchi2024} and the 21 neutron-star candidates of \citet{ElBadry2024NS}. For orbits beyond 100\,pc we require a primary mass of 0.1--1.25\,\Msun, at least 50 XP samples and at least ten \SPHEREx\ channels.

At each fixed companion mass $M_2$, we refit the primary mass $M_1\ge M_2$, age, metallicity and flux scale for a coeval pair. We set $A_V=0$ for GCNS systems within 100\,pc and fit $0\le A_V\le1.5$ for more distant orbits and literature systems, using the extinction curve of \citet{Gordon2023} with $R_V=3.1$. We include a single star and a companion of 0.08\,\Msun. Within 100\,pc we scan 0.10--1.25\,\Msun\ in steps of 0.05\,\Msun. For the other systems we scan 0.15--1.25\,\Msun\ in steps of 0.1\,\Msun. We also test the orbital companion mass and refine the grid near the threshold.

We calculate $q=M_2/M_1$ with the primary mass refitted at each step and select the calibration bin of Appendix~\ref{sec:calibration} using the single-star mass. The upper limit is
\begin{align*}
\Delta(M_2)&=-2\ln\mathcal L(M_2)-\min_{M_2}[-2\ln\mathcal L],\\
M_{2,\rm up}&=\max\{M_2:\Delta(M_2)\le T(q)\}.
\end{align*}
We use $T(q)=K(q)t$ for non-detections and $5.2t_{\rm all}$ for $D\ge25$. For $G<9$ or $M_1>0.9$\,\Msun\ we use the nearest calibration bin. We interpolate linearly at each threshold crossing and take the upper end of the highest allowed interval. For 15 orbits the scan still allows its highest mass, so we report no upper limit. We discard fits that fail to converge, reach a parameter boundary or have $\chi^2/N_{\rm fit}>3$. Outside the GCNS sample we also discard limits below 0.1\,\Msun.

At each $M_2$ we apply the fitted extinction to both model XP spectra and integrate them over the $G$ passband. We take the largest flux ratio over all allowed masses, including the interpolated threshold crossings, as $\beta_{\rm up}$. For 95 per cent of the 1479 systems with GCNS mass-ratio profiles, the two methods agree on whether $\beta_{\rm up}<1$. We compute $\beta_{\rm MS}$ for an MS companion of the orbital mass using the single-star age, metallicity and extinction. When the orbital mass exceeds the primary mass, we set $\beta_{\rm MS}=1$.

For the full triage catalogue we include 2MASS $JHK_s$, allow $0\le A_V\le3$ and use the single-star primary mass without a catalogue prior. We include \SPHEREx\ when at least ten channels are valid, leaving 30\,155 orbits with XP and 2MASS alone. The fit cuts remove 17\,290 orbits, mainly because $\chi^2/N_{\rm fit}>3$ or the flux scale reaches its bound. For the joint fit we use the catalogue observable and its uncertainty,
\[
\mathcal Y_{\rm obs}=(a_0/\varpi)P^{-2/3},\qquad
\chi^2_{\rm orb}=\left[\frac{\mathcal Y_{\rm obs}e^{-u\sigma_s/2}-\mathcal Y}{\sigma_{\mathcal Y}}\right]^2 .
\]
We compute $\mathcal Y=\mathcal A M_1^{1/3}$ from Eq.~\eqref{eq:amrf} at the refitted primary and $\beta_G$. The exponential applies the same parallax adjustment to the spectra and the orbit, with $u$ the flux-scale parameter in units of $\sigma_s$. Setting $\beta_G=0$ at the single-star primary mass gives dark-companion masses with a median ratio of 0.98 to the catalogue values.

We test both orbital roots for the 558 undetected \texttt{Orbital} systems within 100\,pc with $G\ge9$ and $0.3\le M_1/\Msun\le0.9$. We apply the threshold at each root's $q$ and retain faint roots below $q=0.1$. We use 500 draws to propagate the uncertainty in $a_0$ from the orbital significance and a 10 per cent uncertainty in $M_1$. Four systems exclude both roots in at least 95 per cent of the draws. One has no root at the nominal orbit. The other three still exclude both roots when we omit XP samples below 0.45\,\mum\ (Table~\ref{tab:individual}). For \texttt{AstroSpectroSB1} systems we use the mass function and astrometric inclination to constrain the companion mass, allowing the primary mass to vary in the MS companion fit.

\begin{table}
\caption{Systems discussed individually in Appendix~\ref{app:details}.}\label{tab:individual}
\centering
\footnotesize\setlength{\tabcolsep}{3pt}
\begin{tabular}{@{}lp{4.9cm}@{}}
\toprule
System (\Gaia\ DR3 or name) & Outcome \\
\midrule
\multicolumn{2}{l}{\emph{SB2 test of the branch choice}} \\
2080263497110638336 & APOGEE $q=0.50$ rejected by 1627; $q_{\rm XPS,dyn}=0.92$; \texttt{Orbital+M1} range 0.15--0.18 excludes both: possibly a triple \\
1097717538333601408 & branches differ by 1.4 \\
6027732067367229312 & no companion ($D=1.1$); faint root preferred by 4; SB2 $q=0.85$ rejected by 9 \\
\multicolumn{2}{l}{\emph{Substellar candidates}} \\
PM J13580+3141 & substellar companion confirmed by \citet{Barbato2026}; faint root preferred \\
\multicolumn{2}{l}{\emph{Both roots excluded even without XP below 0.45\,\mum}} \\
3467993131525637632 & excluded at 1.5\,$K(q)\,t$; poor single-star fit ($\chi^2/N_{\rm fit}=2.2$); with 2MASS and free $A_V$, one MS companion of 0.18\,\Msun\ fits spectra and orbit better than a dark one by 12 \\
5823631751080095104 & excluded at 3.3\,$K(q)\,t$; $\chi^2/N_{\rm fit}=1.2$; dark, one MS and inner-pair companions within 4 of each other \\
5337468425583762688 & excluded at 1.5\,$K(q)\,t$; $\chi^2/N_{\rm fit}=1.1$; an inner pair of two 0.06\,\Msun\ brown dwarfs fits as well as a dark companion and better than one MS star by 12 \\
\multicolumn{2}{l}{\emph{Compact-companion systems}} \\
J1951$-$0305, J2103$-$4747 & RV-faint \citep{Yamaguchi2024}; MS companion of the published mass not excluded \\
J2102+3703 & neutron-star candidate \citep{ElBadry2024NS}; $\beta_{\rm up}=0.81$ \\
\bottomrule
\end{tabular}
\end{table}

\subsection{Inner pairs}\label{sec:inner-pairs}

We test the inner-pair model on two known eclipsing triples. For \Gaia\ DR3 1748901959855337472 (Fig.~\ref{fig:triple}), a single MS companion gives a photocentre amplitude $3.4\sigma$ below the observed value. Including the masses and fluxes of both inner stars reduces the discrepancy to $1.7\sigma$. For \Gaia\ DR3 6152018658376660992 \citep{ElBadry2026}, we find a total inner-pair mass of 1.30\,\Msun, close to the 1.37\,\Msun\ required by the mass function.

In 26 class-III systems the orbit term favours a luminous companion by more than 25. Their median eccentricity is 0.56, and 8 per cent have nearly circular orbits ($e<0.1$). Among the 472 class-III systems for which we exclude a single MS companion, the median eccentricity is 0.10 and 49 per cent have $e<0.1$.

RV follow-up provides a separate test. For six systems, XPS alone favours a luminous companion ($D>25$), whereas the RVs support a dark companion \citep{ElBadry2026}. When we include the orbit, the dark model gives a lower $-2\ln\mathcal L$ than the luminous model by more than 200 in five of the six systems.

\FloatBarrier
\section{Checking a photocentre orbit with \texttt{spbin}}\label{app:spbin}

We provide tools for testing \Gaia\ photocentre orbits in the public package \texttt{spbin} (\url{https://github.com/jiadonglee/spbin}). Given an orbit and a primary mass, \texttt{spbin.orbit} computes the faint and luminous solutions and predicts the brightening for each. These predictions can be compared with the height of the system above the single-star sequence in the CMD. A metal-poor primary can remain on the sequence even when its companion contributes light (Sect.~\ref{sec:gcns}). For a spectral test, \texttt{spbin} downloads \Gaia\ XP spectra and 2MASS and AllWISE photometry, fits each solution and ranks the fits. If all MS solutions fit the data substantially worse than the dark solution, the spectra favour a companion fainter than an MS star of the orbital mass.

\texttt{spbin} computes $\beta_G$ from the PARSEC $G$ magnitudes of the two components. In this paper we use their model XP spectra. For LP~769-9, both calculations place the luminous root at $q=0.86$ with $\beta_G=0.53$. The NumPy implementation in \texttt{spbin} extends the stellar network to 7500\,K. The model and calibrated threshold $K(q)\,t$ used in this paper are available at Zenodo (see Data availability).

\subsection{White-dwarf companions}\label{app:spbin-wd}

We use \texttt{spbin} to search for light from a white dwarf in six systems with dark companions from \citet{ElBadry2026}. We fit four models to \Gaia\ XP spectra and 2MASS photometry: a single MS star, a coeval MS binary with free mass ratio, a single white dwarf and an MS star with a white dwarf. We use the hydrogen-atmosphere white-dwarf spectra of \citet{Koester2010} at 6--80\,kK and carbon--oxygen cooling tracks up to 1.3\,\Msun. We adopt a Gaussian prior on the white-dwarf mass centred on the dynamical measurement and include far- and near-ultraviolet photometry from the Galaxy Evolution Explorer (GALEX) where available \citep{Martin2005}.

With XP and 2MASS alone, we cannot reliably identify a white dwarf in any of the six systems. Where the optical fit favours a white dwarf, its temperature reaches the upper grid limit of 80\,kK. For \Gaia\ DR3 2919995917769953408 (dynamical companion mass $1.26\pm0.07$\,\Msun, $D=0.9$), we favour an MS star with a white dwarf only when we include the GALEX far-ultraviolet flux. The fitted white dwarf has $\teff\approx24$\,kK and contributes 0.4 per cent of the total $G$-band light.

\FloatBarrier
\end{appendix}
\end{document}